\def\year{2021}\relax
\documentclass[letterpaper]{article} 
\usepackage{aaai21}  
\usepackage{times}  
\usepackage{helvet} 
\usepackage{courier}  
\usepackage[hyphens]{url}  
\usepackage{graphicx} 
\usepackage{natbib}  
\usepackage{caption} 
\usepackage{threeparttable}
\usepackage{booktabs}
\usepackage{multirow}
\usepackage{subcaption}
\usepackage{enumitem}
\usepackage{amsmath}
\usepackage{amssymb}
\usepackage[hang,flushmargin]{footmisc}
\newcommand{\Lagr}{\mathcal{L}}

\title{Learning to Enhance Visual Quality via Hyperspectral Domain Mapping}
\author{
    Harsh Sinha, \textsuperscript{\rm 1}
    Aditya Mehta, \textsuperscript{\rm 1}
    Murari Mandal, \textsuperscript{\rm 2}
    Pratik Narang \textsuperscript{\rm 1}\\
}
\affiliations{
    \textsuperscript{\rm 1}Department of Computer Science and Information Systems, BITS Pilani, Rajasthan, India, 333031\\
    \textsuperscript{\rm 2}Department of Computer Science and Engineering, IIIT Kota, India\\
    \{h20130838, h20150808, pratik.narang\}@pilani.bits-pilani.ac.in,
    murarimandal.cv@gmail.com \\
}
\begin{document}
\maketitle
\begin{abstract}
Deep learning based methods have achieved remarkable success in image restoration and enhancement, but most such methods rely on RGB input images. These methods fail to take into account the rich spectral distribution of natural images. We propose a deep architecture, \textsc{SpecNet}, which computes spectral profile to estimate pixel-wise dynamic range adjustment of a given image. First, we employ an unpaired cycle-consistent framework to generate hyperspectral images (HSI) from low-light input images. HSI is further used to generate a normal light image of the same scene. We incorporate a self-supervision and a spectral profile regularization network to infer a plausible HSI from an RGB image. We evaluate the benefits of optimizing the spectral profile for real and fake images in low-light conditions on the LOL Dataset.
\end{abstract}

\section{Introduction}
Human visual perception is acquainted with high-contrast images that are characterized by high contrast, good visibility, and minimal noise. Thus researchers have focused extensively on developing computer-vision techniques to improve the visual perception of images. Such algorithms have broad applicability, such as all-weather autonomous vehicles and illumination-invariant face detection. 

Low-light image enhancement is a well-studied problem, and researchers have proposed several methods to address this problem. These methods include histogram equalization, dehazing-based approaches, and retinex theory. Although these representative state-of-the-art methods produce good results, they are limited in terms of model capacity for illumination and reflectance decomposition. Such constraints are hand-crafted and require careful hyperparameter-optimization. To mitigate this problem, researchers have used CNNs for low-level image processing. Owing to the extensive success of GANs for the problem of image-to-image translation, we build a framework that can generate visually-pleasing images through spectral guidance.

In this paper, we propose \textsc{SpecNet} which optimizes a spectral profile to achieve superior results. We first use a cycle-consistent framework to reconstruct hyperspectral images from RGB images which is further used to restore proper illumination for the given low-light or dark image. The primary GAN framework used for hyperspectral reconstruction has been carefully modified to incorporate a spectral-profile optimization framework, ultimately aimed at generating visually-pleasing images. Finally, we perform extensive set of experiments to evaluate the effectiveness of the model.
\begin{figure}[t!]
 \centering
     \begin{subfigure}[t]{0.23\columnwidth}
             \includegraphics[width=\columnwidth,height=22mm]{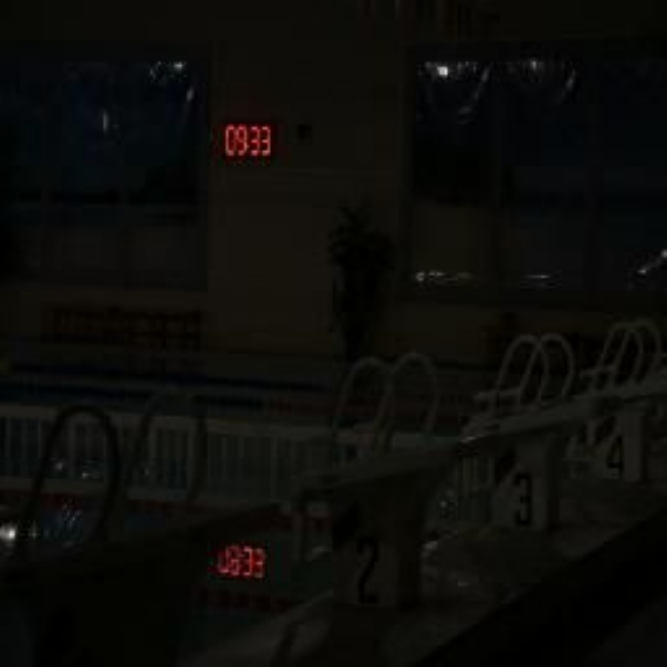}
             \caption{Dark Image}
             \label{subfig:intro_hazy}
     \end{subfigure}
     \begin{subfigure}[t]{0.23\columnwidth}
             \includegraphics[width=\columnwidth,height=22mm]{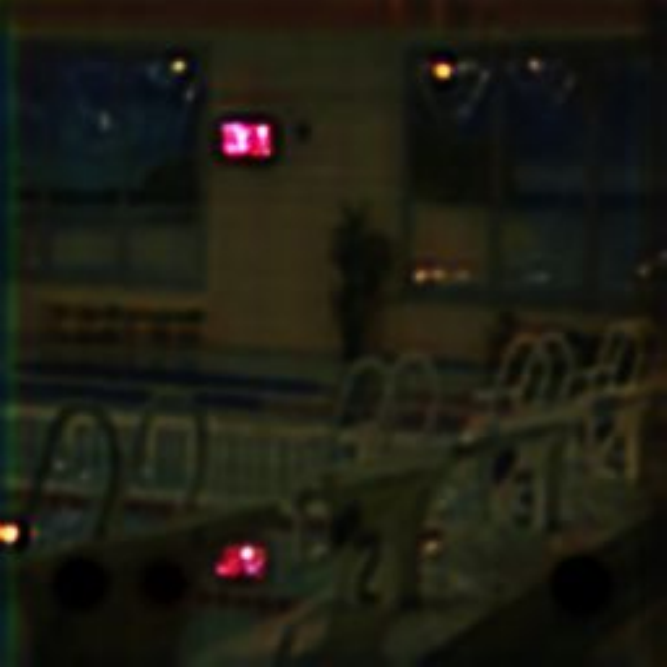}
             \caption{Reconstructed HSI}
             \label{subfig:intro_hsi}
     \end{subfigure}
     \begin{subfigure}[t]{0.23\columnwidth}            
             \includegraphics[width=\columnwidth,height=22mm]{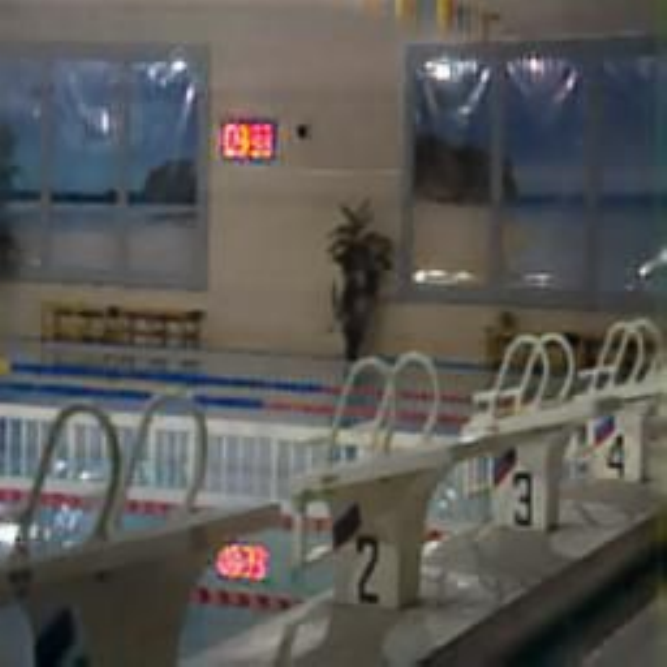}
             \caption{Enh. Output}
             \label{subfig:intro_output}
     \end{subfigure}
      \begin{subfigure}[t]{0.23\columnwidth}
             \includegraphics[width=\columnwidth,height=22mm]{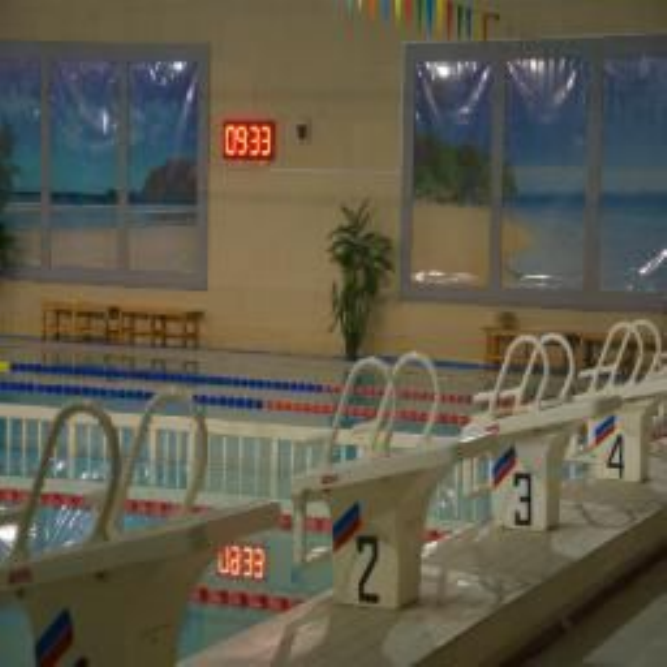}
             \caption{Ground Truth}
             \label{subfig:intro_gt}
     \end{subfigure}%
     \caption{A sample dark image along with the reconstructed HSI and the output obtained using \textsc{SpecNet}. 
     }
     \label{fig:intro}
 \end{figure}
 
 \begin{figure}
    \centering
    \includegraphics[width=\columnwidth, height=35mm]{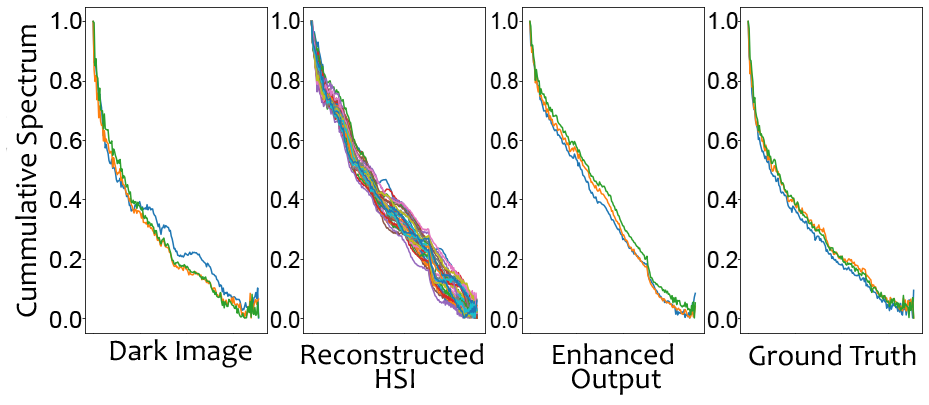}
    \caption{Multi-channel cumulative spectral profile of  sample dark image along with the reconstructed HSI and the output obtained using \textsc{SpecNet}}
    \label{fig:my_label}
\end{figure}
\begin{figure*}
    \centering
    \begin{subfigure}[t]{0.13\textwidth}
    \caption{Dark Image}
    \includegraphics[width=\textwidth,height=20mm]{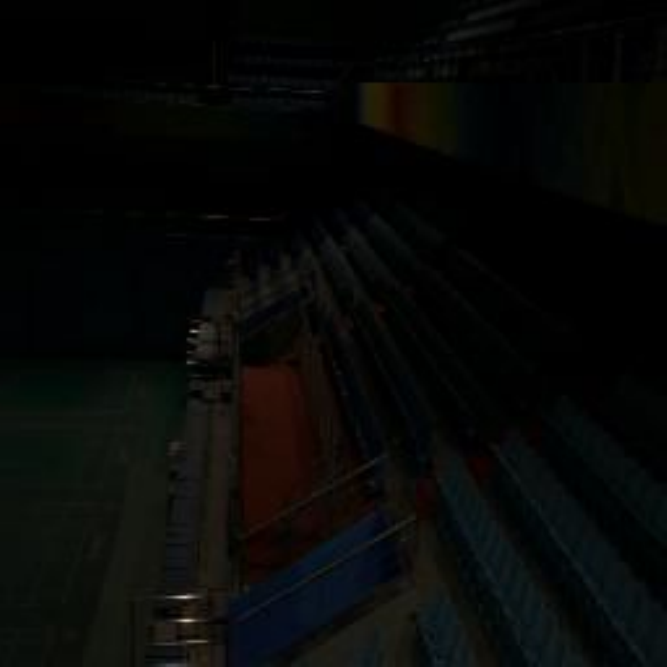}
    \end{subfigure}
    \begin{subfigure}[t]{0.13\textwidth}
    \caption{U-Net}
    \includegraphics[width=\textwidth,height=20mm]{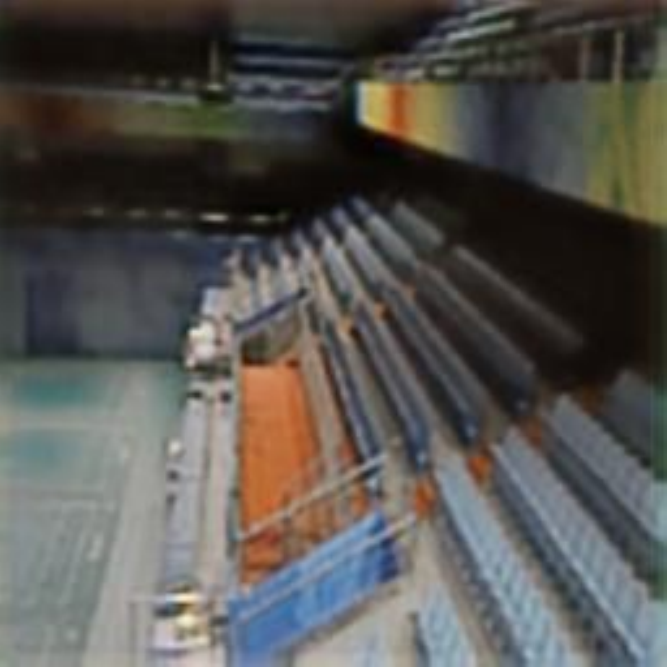}
    \end{subfigure}
    \begin{subfigure}[t]{0.13\textwidth}
    \caption{Pix2Pix}
    \includegraphics[width=\textwidth,height=20mm]{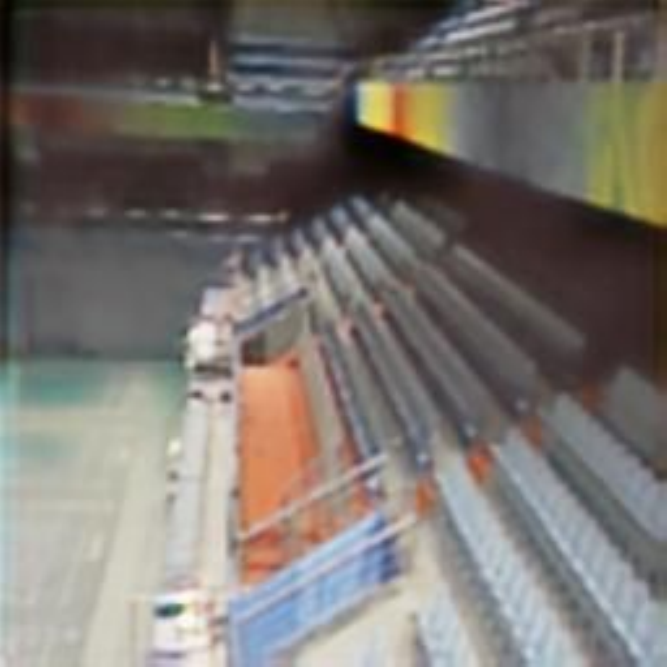}
    \end{subfigure}
    \begin{subfigure}[t]{0.13\textwidth}
    \caption{CycleGAN}
    \includegraphics[width=\textwidth,height=20mm]{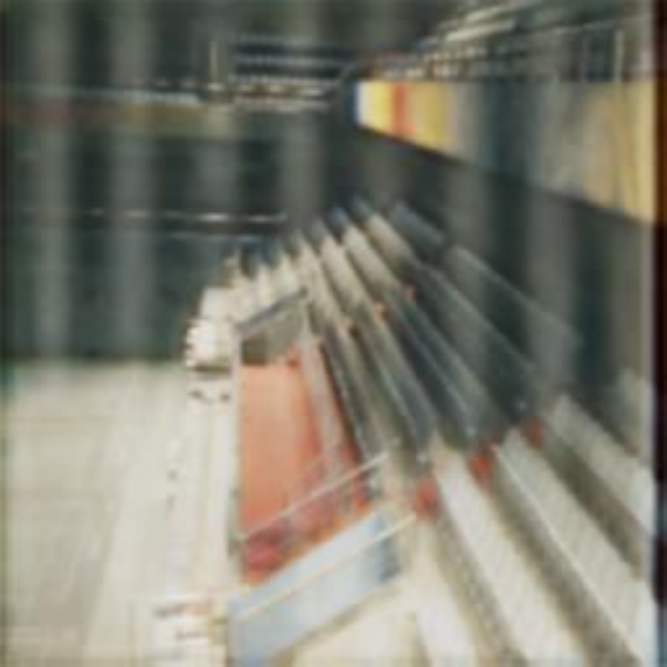}
    \end{subfigure}
    \begin{subfigure}[t]{0.13\textwidth}
    \caption{EnlightenGAN}
    \includegraphics[width=\textwidth,height=20mm]{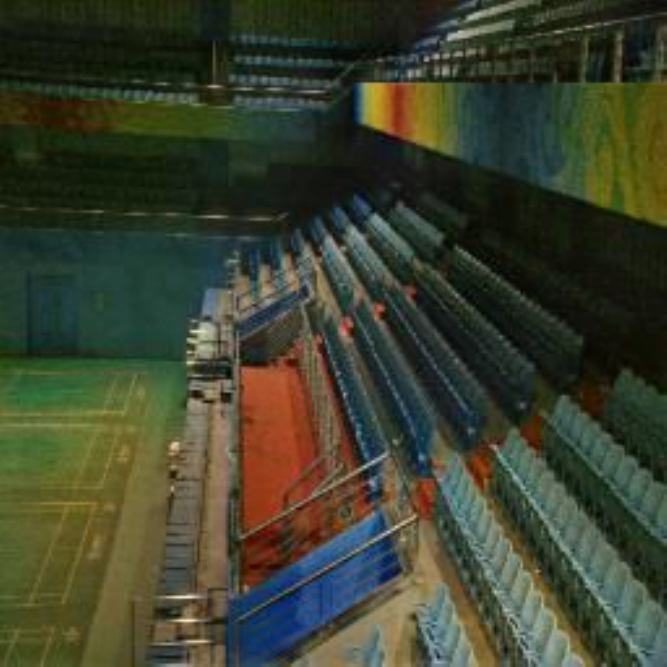}
    \end{subfigure}
    \begin{subfigure}[t]{0.13\textwidth}
    \caption{\textbf{SpecNet}}
    \includegraphics[width=\textwidth,height=20mm]{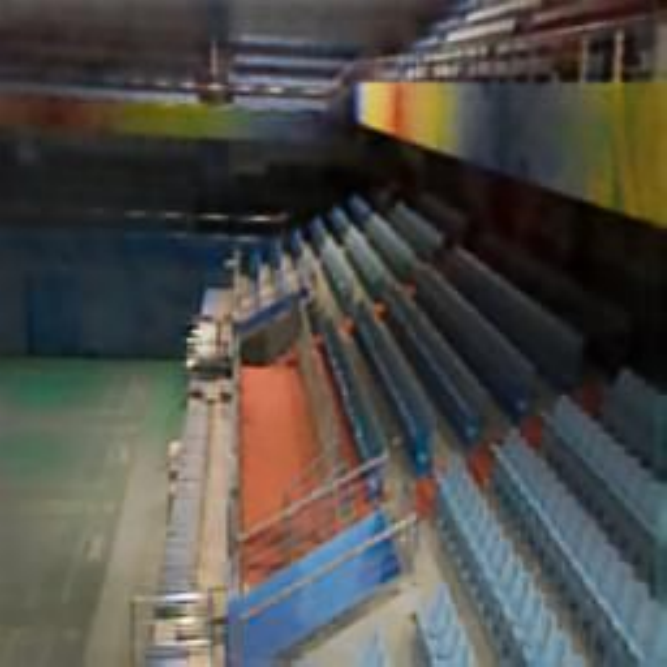}
    \end{subfigure}
    \begin{subfigure}[t]{0.13\textwidth}
    \caption{Ground Truth}
    \includegraphics[width=\textwidth,height=20mm]{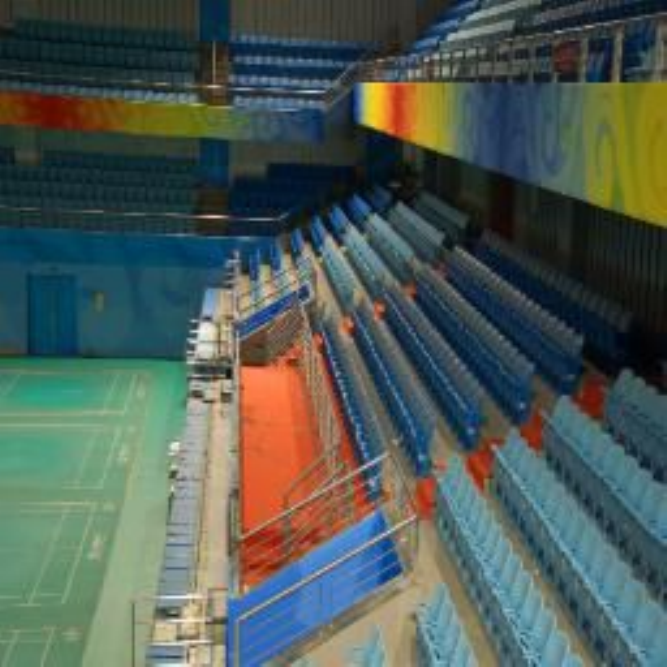}
    \end{subfigure} 
    
    
    \caption{Qualitative comparison for different models as described in Table \ref{table:results}.} 
    \label{fig:ablation_comparison}

\end{figure*}

\section{Proposed Method}
To propose \textsc{SpecNet}, we hypothesize that multi-band information in the reconstructed hyperspectral images can improve the perceptual quality of images. First of all, we create a spanned 31-channel RGB image matrix to imitate the 31-channel HSI, to ease the under-constrained problem of HSI reconstruction from RGB images. The framework can be viewed as a cascaded GAN approach. The first GAN takes an unsupervised cycle-consistent approach to reconstruct HSI, which is fed into another cGAN to generate the final enhanced output image. To solve the under-constrained problem of HSI reconstruction we make use of several guiding principles such as task-aided supervision and spectral-profile optimization.
\vspace{-0.5em}
\paragraph{Aided Supervision} The lack of large-scale hyperspectral image datasets poses a problem in learning an output distribution that can imitate the underlying original hyperspectral values. A task-aided supervision addresses this distributional discrepancy. We use our original low-light enhancement task as an auxiliary optimization task to aid hyperspectral reconstruction. We modify the cycle-consistency loss as

\begin{equation}
\label{eq1}
\begin{split}
\Lagr_{cyc} =& \Arrowvert y-G_h(G_x(x)) \Arrowvert_2^2 + \Arrowvert h-G_x(G_h(h)) \Arrowvert_2^2. 
\end{split}
\end{equation}
where $(x, y)$ refers to dark and ehanced RGB images respectively, $G_x, G_h$ refer to dual generators used for cycle-consistency and $h$ refers to HSI. 

\vspace{-0.5em}
\paragraph{Spectral-profile Optimization} As the primary task of the framework is to produce enhanced images, we incorporate a network to generate spectral-profile using multi-channel power spectrum from 2D Fourier transform \cite{durall2020watch}. The network was used to regularize the spectral distribution of reconstructed HSI. The motivation is to induce alignment in spectral distributional discrepancy in the reconstructed HSI. This is achieved by jointly optimizing the algorithm with a spectral-profile generator that discriminates between spectral profiles of reconstructed HSI and real RGB images. By minimizing the mean squared error, the algorithm encourages spectrally-enduring HSI.
\vspace{-0.5em}
\paragraph{Multi-layer Colorization Space} The multi-layer colorization space is constructed using different color models such as HSV, YCrCb, and LAB concatenated together with RGB which results in a 12-channel input image \cite{mehta2020domainaware}. This is fed into cGAN along with the reconstructed HSI.

\section{Experimental Evaluation}
The experimental results in terms of PSNR and SSIM on LOL dataset \cite{wei2018deep} are compiled in Table   \ref{table:results}. \textsc{SpecNet} outperforms the existing state-of-the-art techniques in terms of PSNR and SSIM.

\begin{table}
    \centering
    \begin{threeparttable}
    \begin{tabular}{c|l l}
    \toprule
    Method & \multicolumn{1}{c}{SSIM} & \multicolumn{1}{c}{PSNR} \\ 
    \midrule
    \midrule
    
    U-Net & 0.7397  & 21.500  \\ 
    Pix2Pix & 0.7307 & 20.483    \\ 
    CycleGAN & 0.6850 & 20.348   \\
    EnlightenGAN & 0.7694 & 23.202  \\
    \textbf{\textsc{SpecNet}} & \textbf{0.8052} & \textbf{22.330}  \\ 
    \bottomrule
    \end{tabular}
    \end{threeparttable}
    \caption{Comparative results on LOL dataset}
    \label{table:results}
\end{table}
The proposed \textsc{SpecNet} consists of several components which add to performance through cumulative effort.To delineate the contributions of different components, several models were trained apart from the final model. The comparative performance is summarized in Table \ref{table:ablation_study}.
    

\begin{table}
\resizebox{\columnwidth}{!}{
    \begin{tabular}{c|c c |c}
    \toprule
        &  \multicolumn{2}{c|}{Components} & \\ [4pt]
        \cline{2-3}
         Method & \begin{tabular}[c]{@{}c@{}}Spectral Profile \\ Optimization\end{tabular} & \begin{tabular}[c]{@{}c@{}}Multi-Layer\\ Colorization Space\end{tabular} & \begin{tabular}[c]{@{}c@{}} SSIM \end{tabular} \\ \midrule \midrule
        Model-1 & & & 0.6784 \\
        Model-2 & \checkmark & & 0.7244  \\
        \textbf{\textsc{SpecNet}} & \checkmark & \checkmark & \textbf{0.8052} \\
        \bottomrule
    \end{tabular}
    }
    \caption{Ablation Models}
    \label{table:ablation_study}
\end{table}
\section{Conclusions}
This work demonstrates the use of spectral-profile optimization for low-light image enhancement using a cascaded GAN framework, referred to as \textsc{SpecNet}. It reconstructs HSI from low-light RGB images and an enhanced cGAN generates enhanced output images using reconstructed hyperspectral images. The model utilizes color spaces by concatenating a 12-channel multi-layer color space with the reconstructed HSI. Further, an ablation study is conducted which substantiates the contribution of individual components in the framework.

\begin{figure*}
    \centering
    \begin{subfigure}[t]{0.13\textwidth}
    \caption{Dark Image\\\phantom{~\shortcite{wei2018deep}}}
    \includegraphics[width=\textwidth,height=25mm]{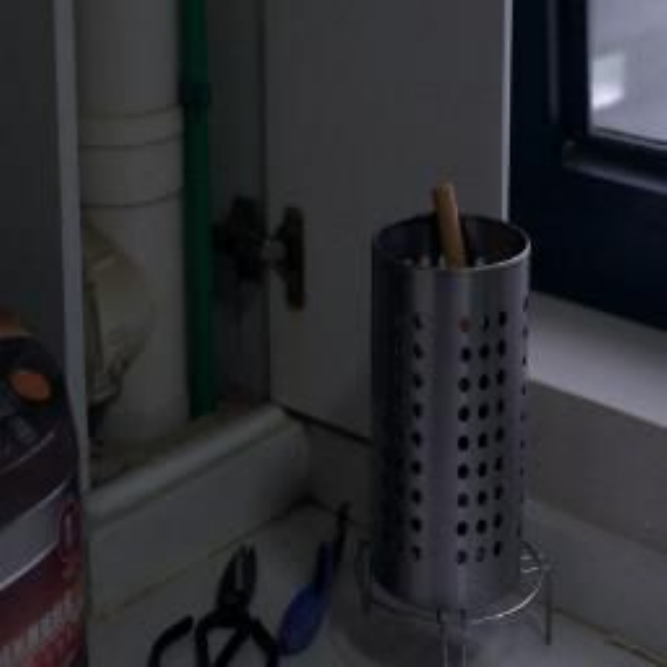}
    \end{subfigure}
    \begin{subfigure}[t]{0.13\textwidth}
    \caption{U-Net \\\shortcite{ronneberger2015u}}
    \includegraphics[width=\textwidth,height=25mm]{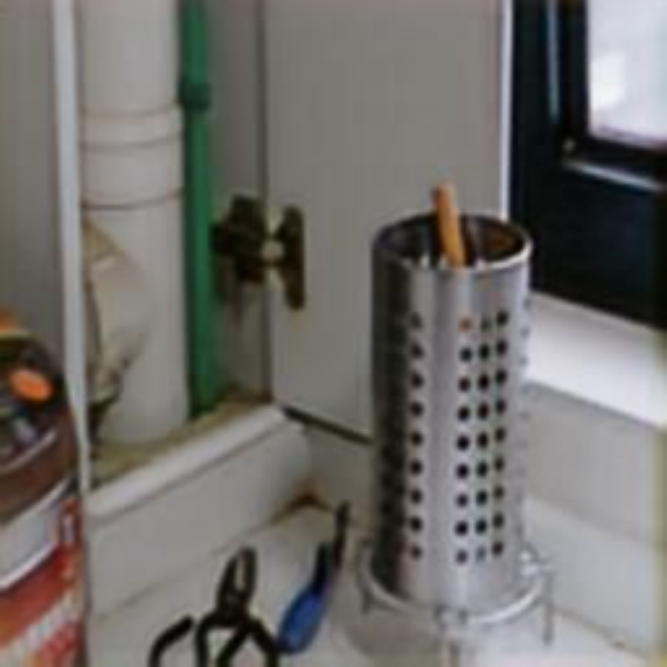}
    \end{subfigure}
    \begin{subfigure}[t]{0.13\textwidth}
    \caption{Pix2Pix\\\shortcite{isola2017image}}
    \includegraphics[width=\textwidth,height=25mm]{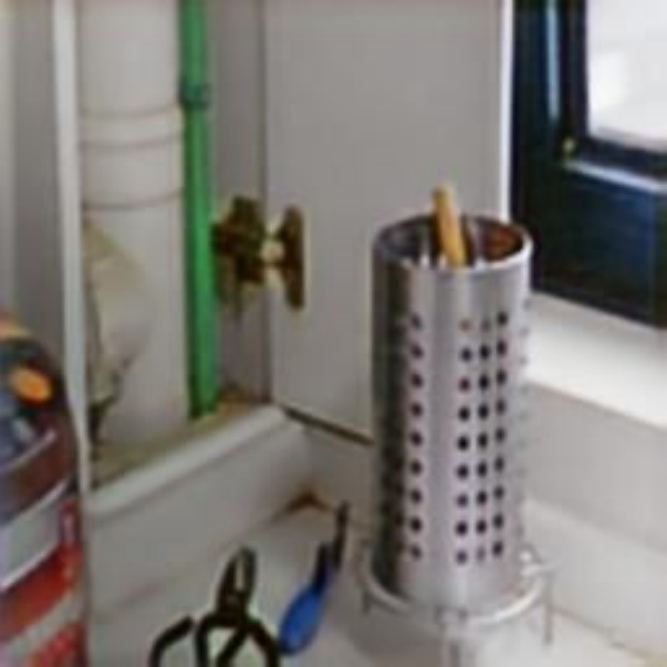}
    \end{subfigure}
    \begin{subfigure}[t]{0.13\textwidth}
    \caption{EnlightenGAN\\  \shortcite{jiang2019enlightengan}}
    \includegraphics[width=\textwidth,height=25mm]{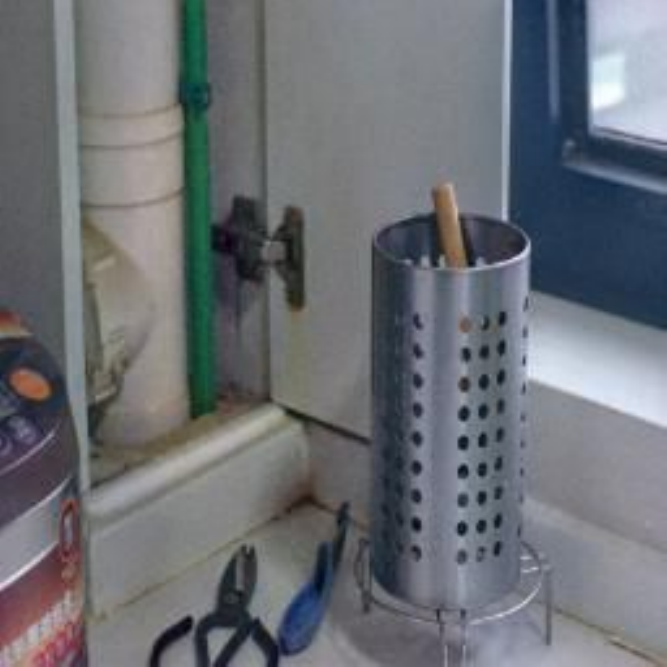}
    \end{subfigure}
    \begin{subfigure}[t]{0.13\textwidth}
    \caption{CycleGAN\\\shortcite{zhu2017unpaired}}
    \includegraphics[width=\textwidth,height=25mm]{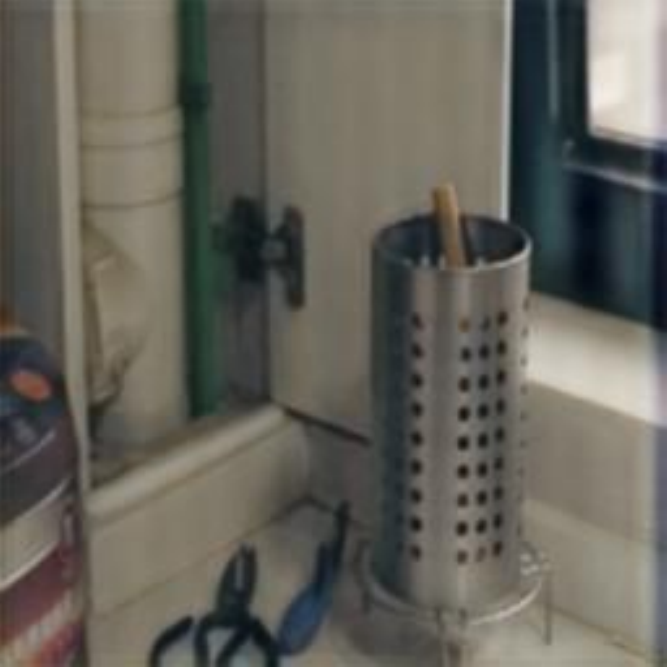}
    \end{subfigure}
    \begin{subfigure}[t]{0.13\textwidth}
    \caption{\textbf{SpecNet}\\ \phantom{Ours}}
    \includegraphics[width=\textwidth,height=25mm]{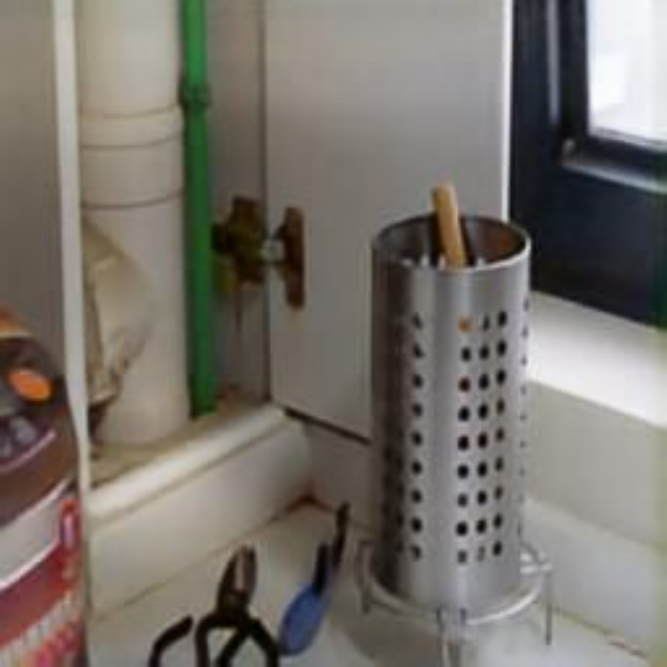}
    \end{subfigure}
    \begin{subfigure}[t]{0.13\textwidth}
    \caption{Ground Truth\\ \phantom{\shortcite{wei2018deep}}}
    \includegraphics[width=\textwidth,height=25mm]{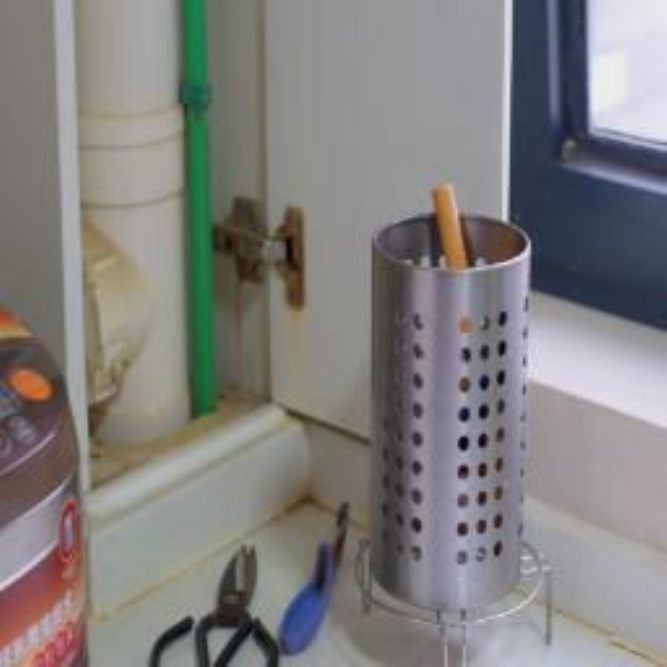}
    \end{subfigure} 
    \\
    \begin{subfigure}[t]{0.13\textwidth}
    \includegraphics[width=\textwidth,height=25mm]{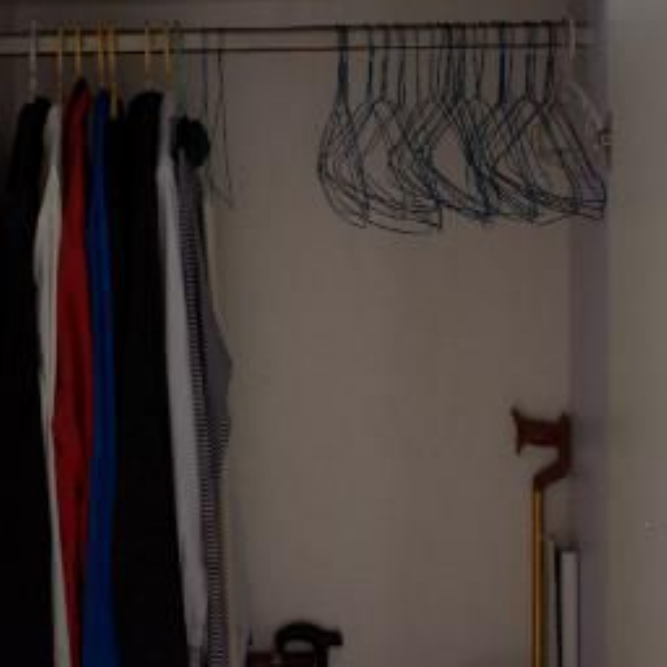}
    \end{subfigure}
    \begin{subfigure}[t]{0.13\textwidth}
    \includegraphics[width=\textwidth,height=25mm]{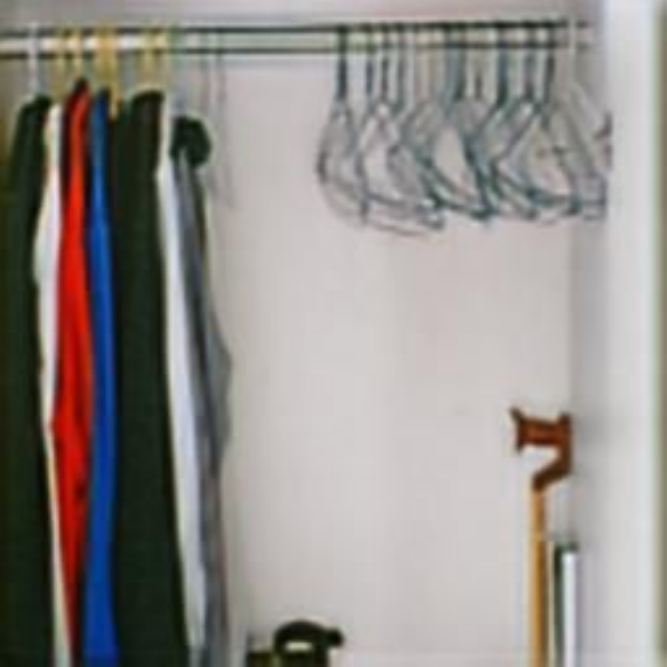}
    \end{subfigure}
    \begin{subfigure}[t]{0.13\textwidth}
    \includegraphics[width=\textwidth,height=25mm]{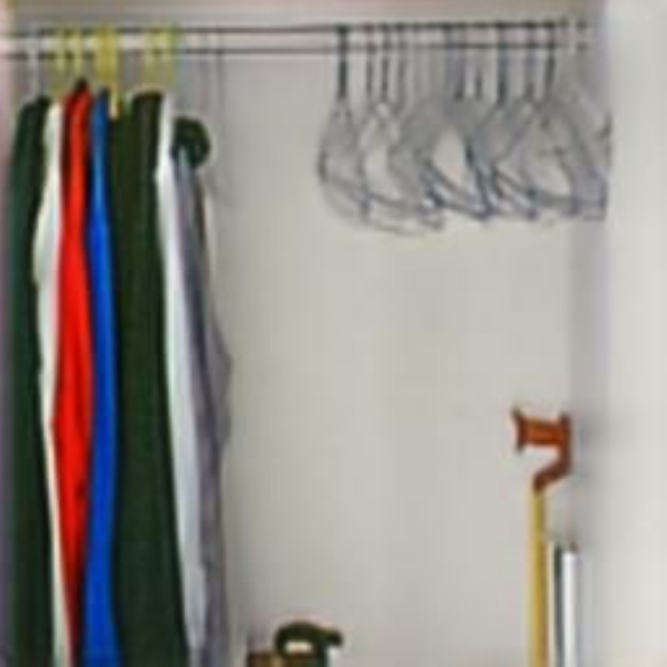}
    \end{subfigure}
    \begin{subfigure}[t]{0.13\textwidth}
    \includegraphics[width=\textwidth,height=25mm]{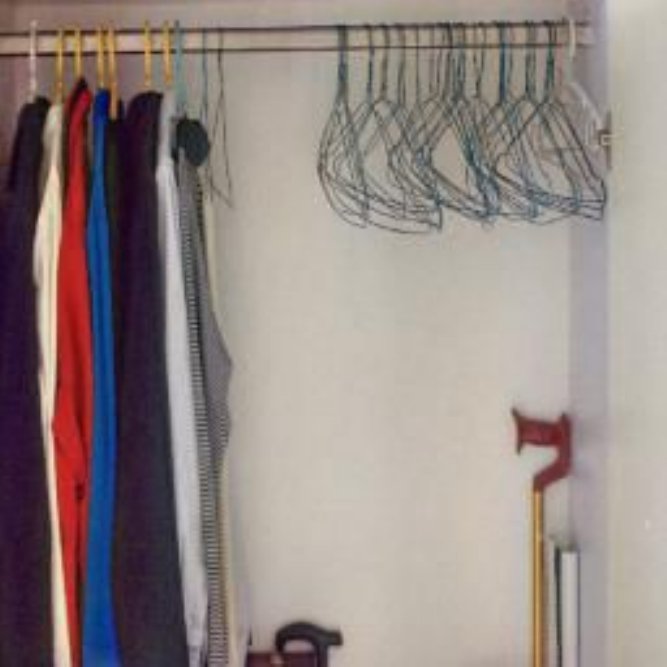}
    \end{subfigure}
    \begin{subfigure}[t]{0.13\textwidth}
    \includegraphics[width=\textwidth,height=25mm]{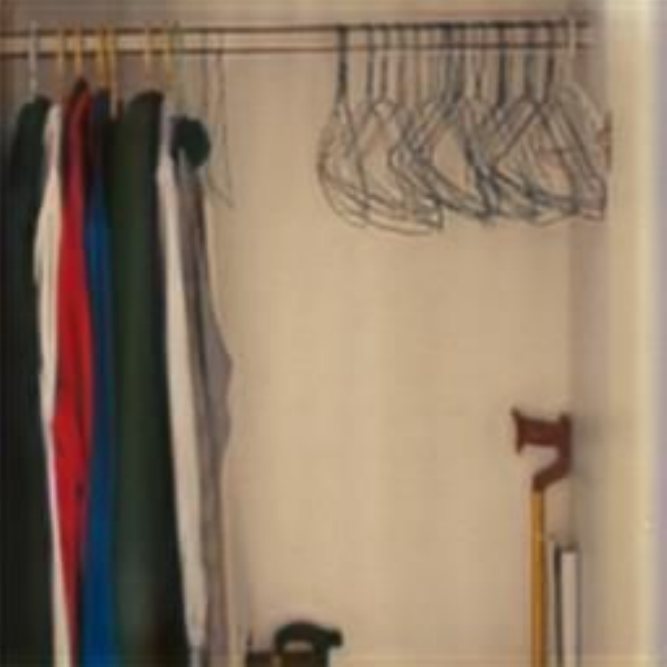}
    \end{subfigure}
    \begin{subfigure}[t]{0.13\textwidth}
    \includegraphics[width=\textwidth,height=25mm]{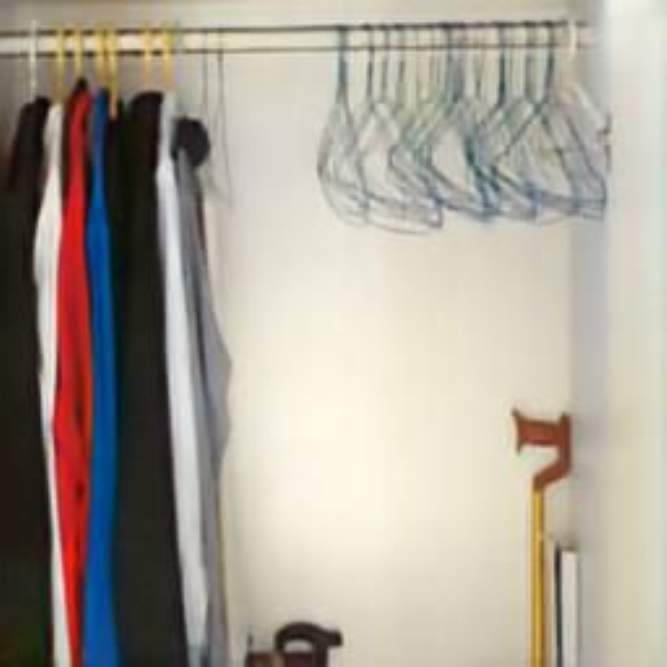}
    \end{subfigure}
    \begin{subfigure}[t]{0.13\textwidth}
    \includegraphics[width=\textwidth,height=25mm]{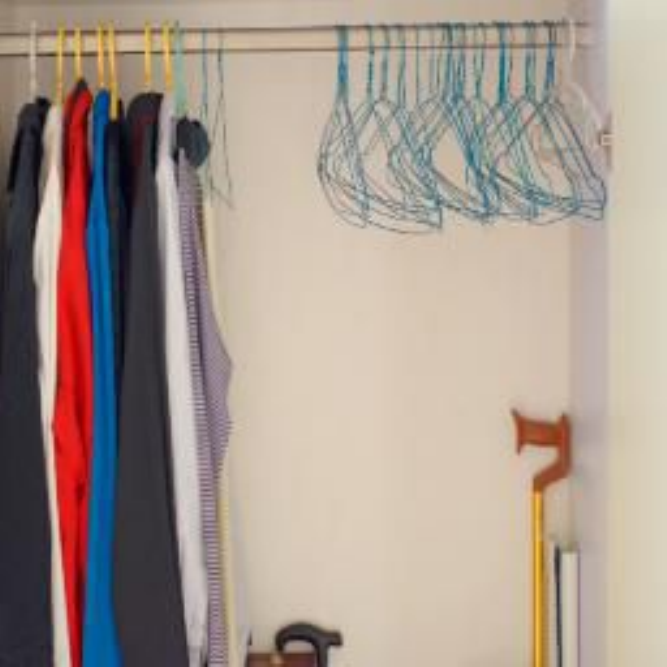}
    \end{subfigure} 
    \\
    \begin{subfigure}[t]{0.13\textwidth}
    \includegraphics[width=\textwidth,height=25mm]{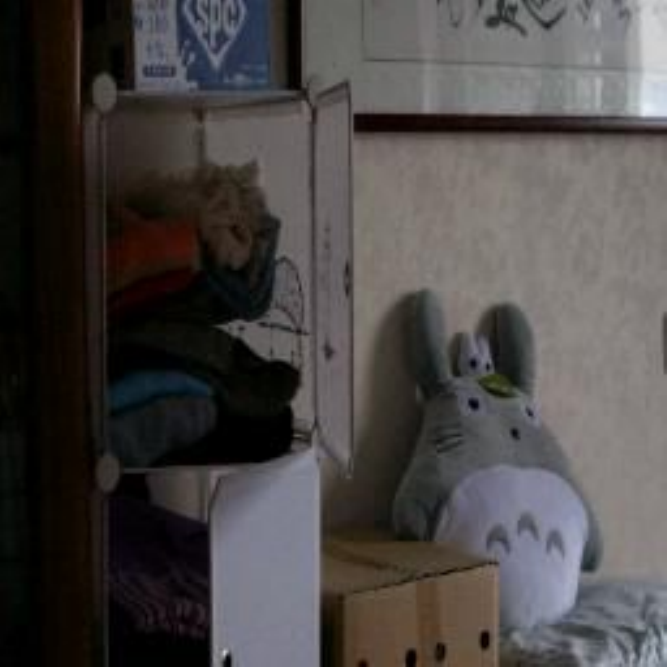}
    \end{subfigure}
    \begin{subfigure}[t]{0.13\textwidth}
    \includegraphics[width=\textwidth,height=25mm]{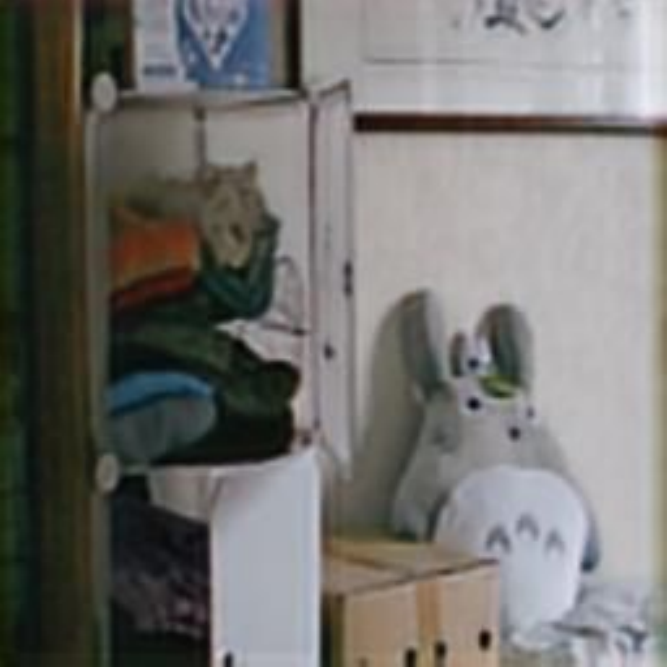}
    \end{subfigure}
    \begin{subfigure}[t]{0.13\textwidth}
    \includegraphics[width=\textwidth,height=25mm]{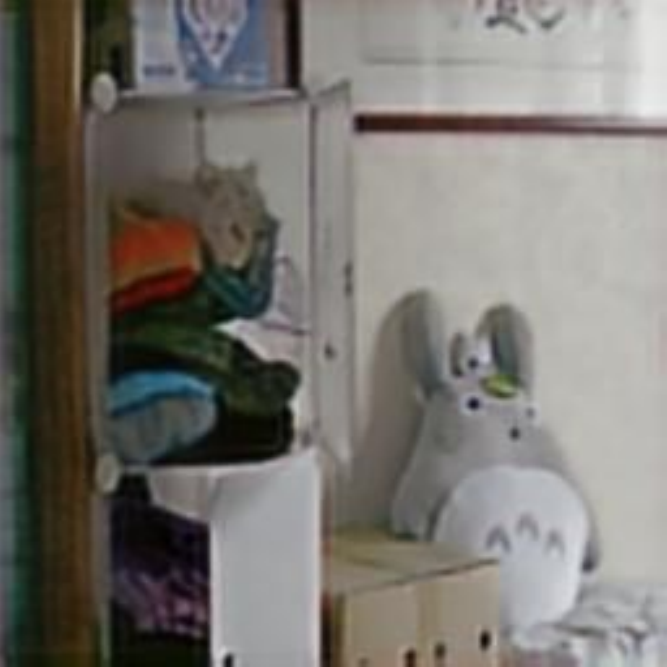}
    \end{subfigure}
    \begin{subfigure}[t]{0.13\textwidth}
    \includegraphics[width=\textwidth,height=25mm]{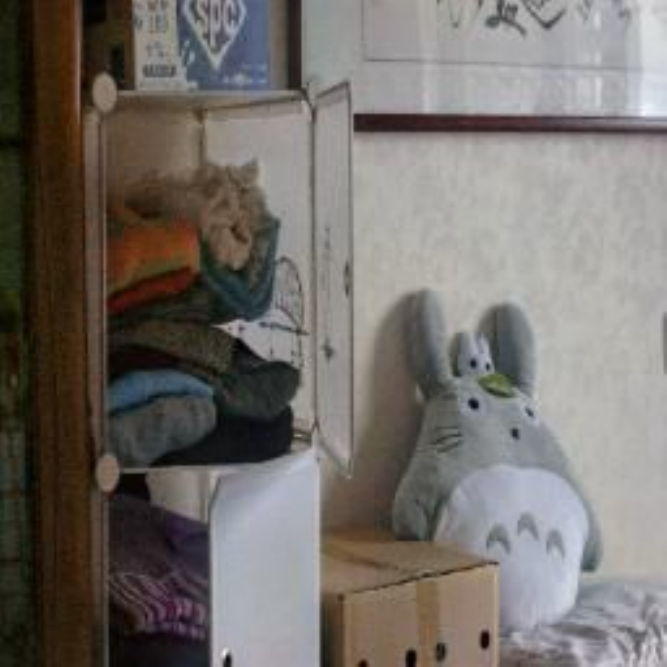}
    \end{subfigure}
    \begin{subfigure}[t]{0.13\textwidth}
    \includegraphics[width=\textwidth,height=25mm]{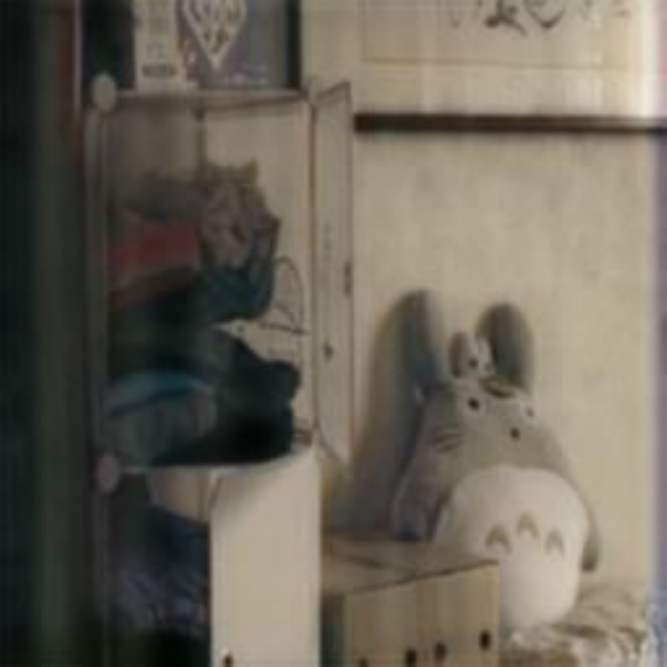}
    \end{subfigure}
    \begin{subfigure}[t]{0.13\textwidth}
    \includegraphics[width=\textwidth,height=25mm]{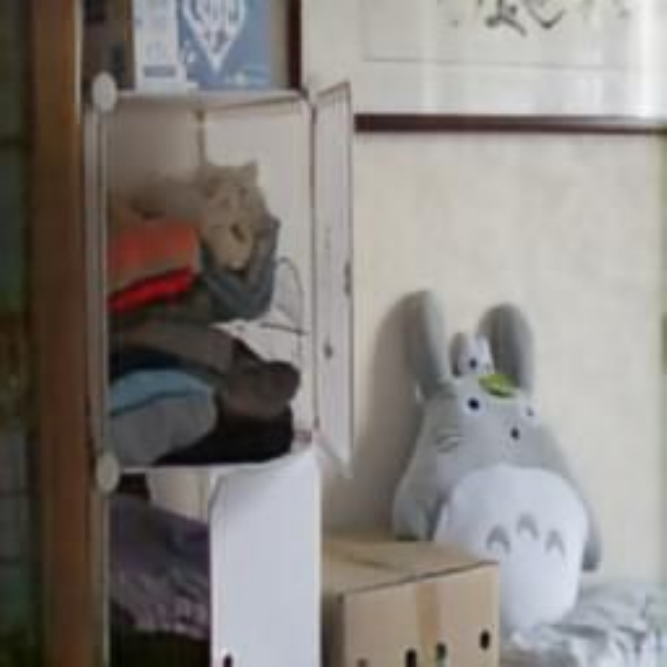}
    \end{subfigure}
    \begin{subfigure}[t]{0.13\textwidth}
    \includegraphics[width=\textwidth,height=25mm]{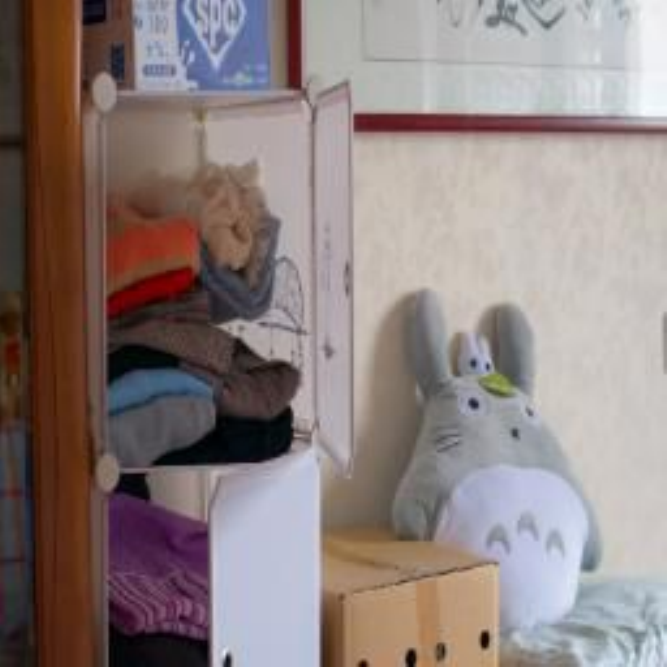}
    \end{subfigure} 
    \\
    \begin{subfigure}[t]{0.13\textwidth}
    \includegraphics[width=\textwidth,height=25mm]{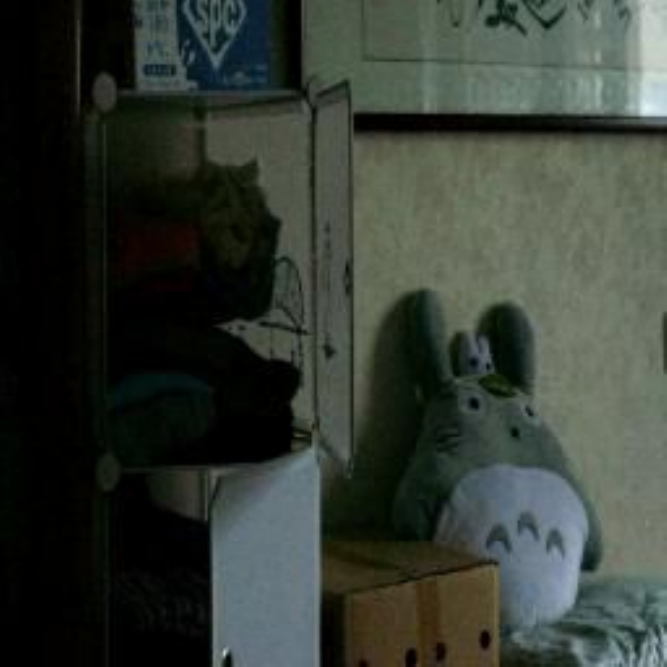}
    \end{subfigure}
    \begin{subfigure}[t]{0.13\textwidth}
    \includegraphics[width=\textwidth,height=25mm]{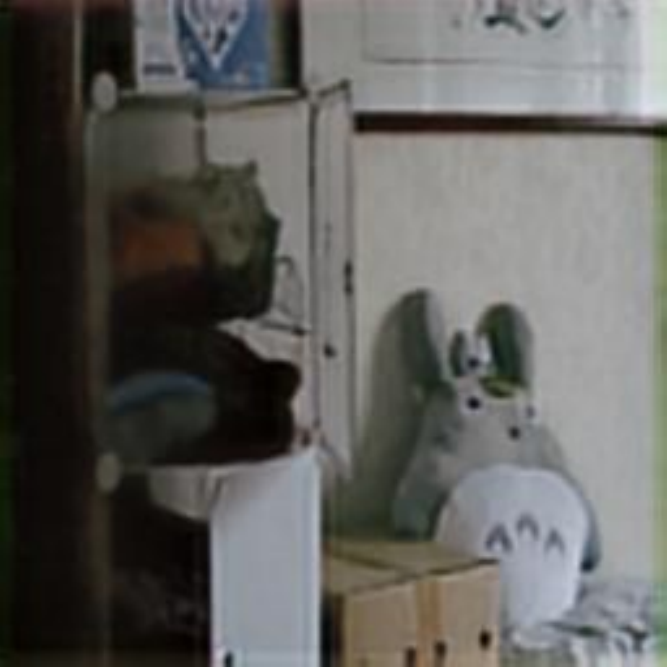}
    \end{subfigure}
    \begin{subfigure}[t]{0.13\textwidth}
    \includegraphics[width=\textwidth,height=25mm]{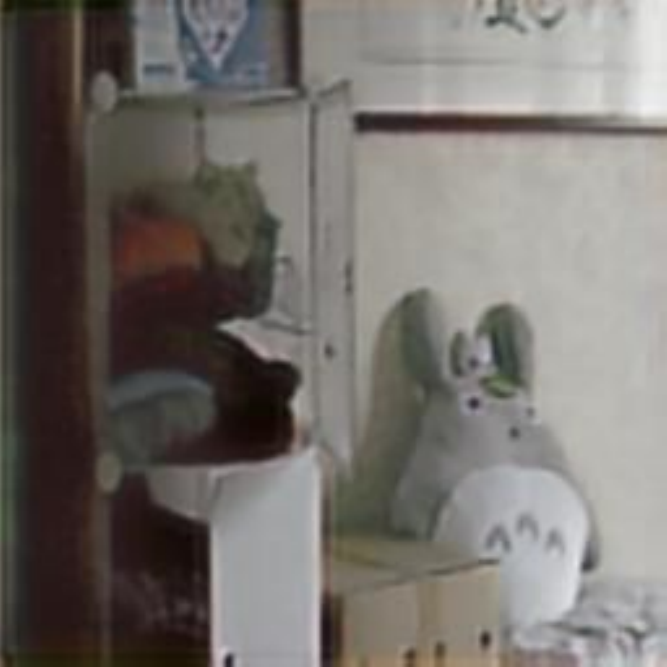}
    \end{subfigure}
    \begin{subfigure}[t]{0.13\textwidth}
    \includegraphics[width=\textwidth,height=25mm]{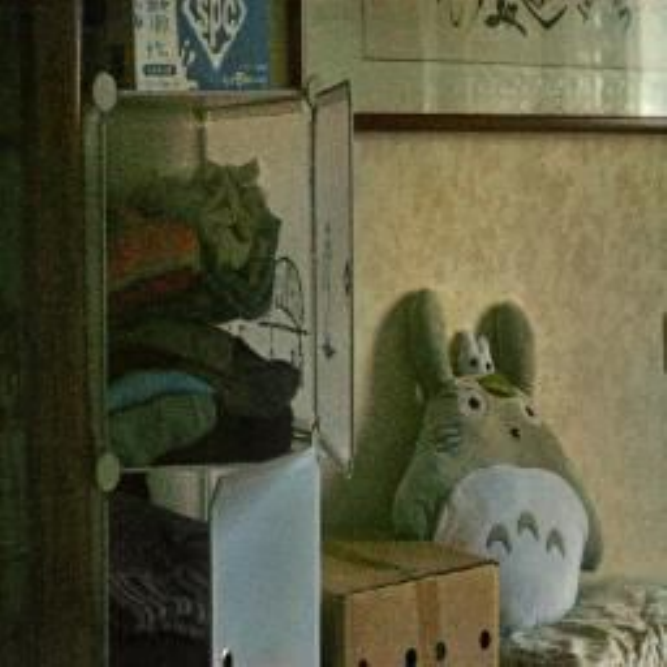}
    \end{subfigure}
    \begin{subfigure}[t]{0.13\textwidth}
    \includegraphics[width=\textwidth,height=25mm]{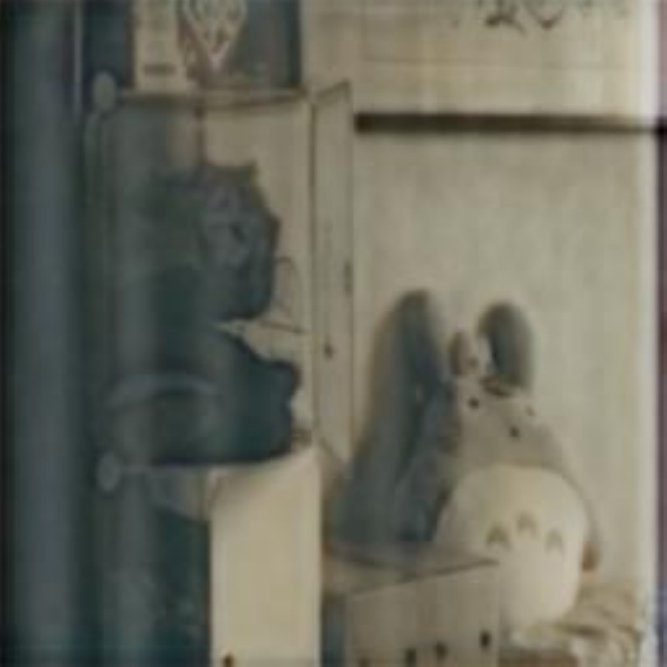}
    \end{subfigure}
    \begin{subfigure}[t]{0.13\textwidth}
    \includegraphics[width=\textwidth,height=25mm]{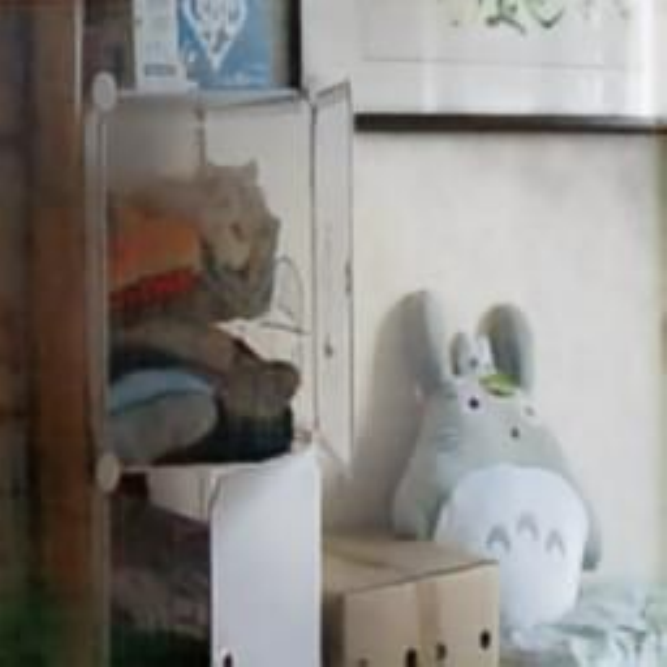}
    \end{subfigure}
    \begin{subfigure}[t]{0.13\textwidth}
    \includegraphics[width=\textwidth,height=22mm]{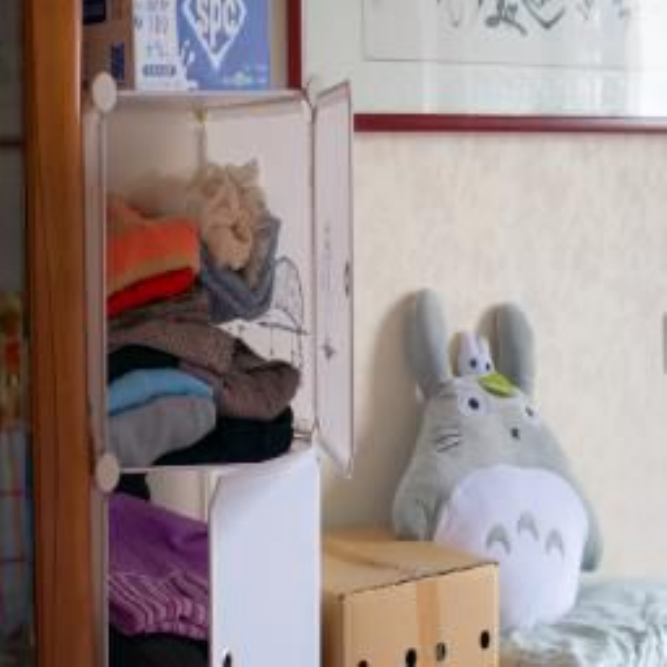}
    \end{subfigure} 
    \\
    \begin{subfigure}[t]{0.13\textwidth}
    \includegraphics[width=\textwidth,height=25mm]{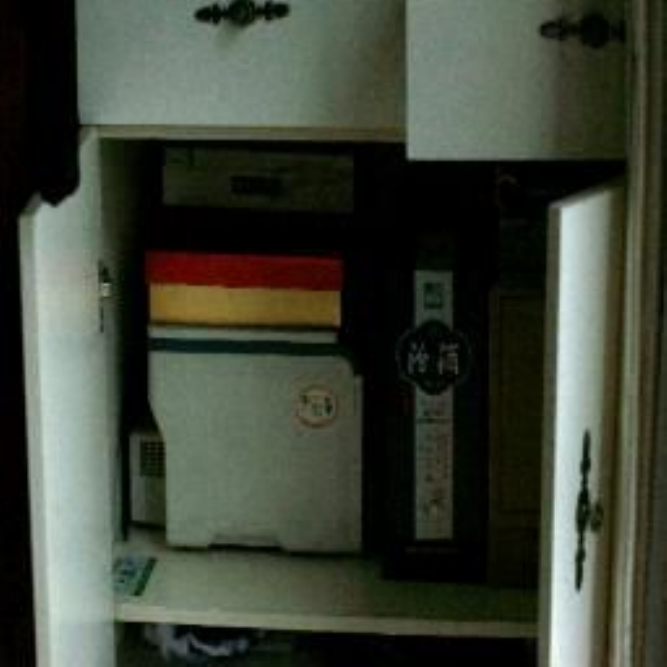}
    \end{subfigure}
    \begin{subfigure}[t]{0.13\textwidth}
    \includegraphics[width=\textwidth,height=25mm]{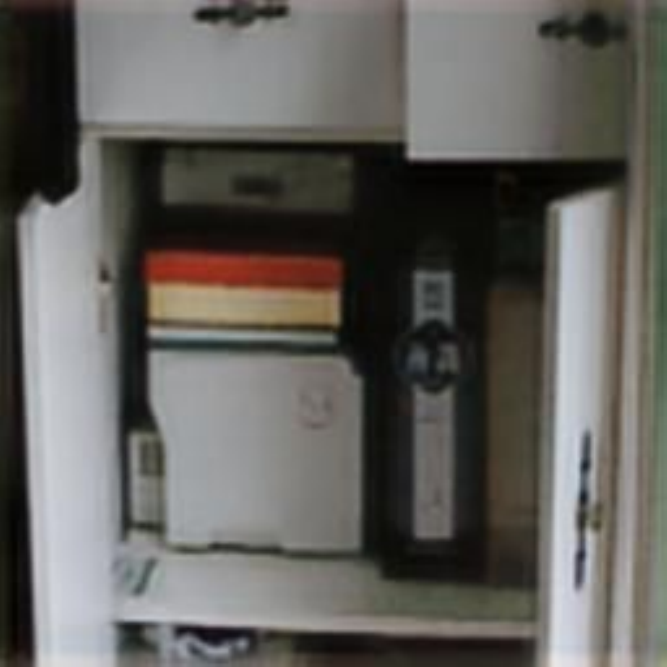}
    \end{subfigure}
    \begin{subfigure}[t]{0.13\textwidth}
    \includegraphics[width=\textwidth,height=25mm]{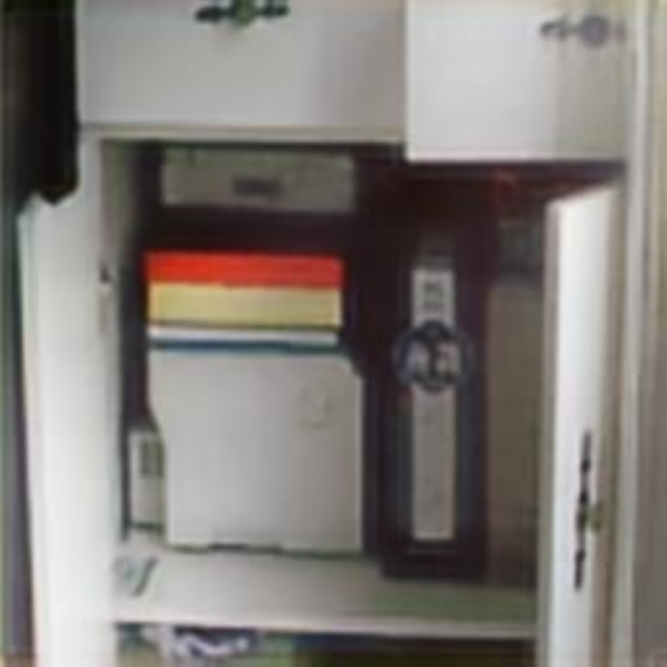}
    \end{subfigure}
    \begin{subfigure}[t]{0.13\textwidth}
    \includegraphics[width=\textwidth,height=25mm]{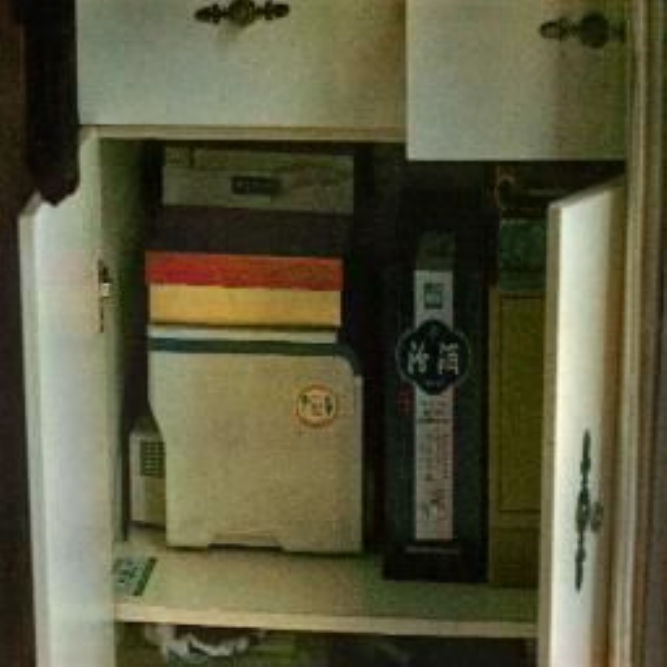}
    \end{subfigure}
    \begin{subfigure}[t]{0.13\textwidth}
    \includegraphics[width=\textwidth,height=25mm]{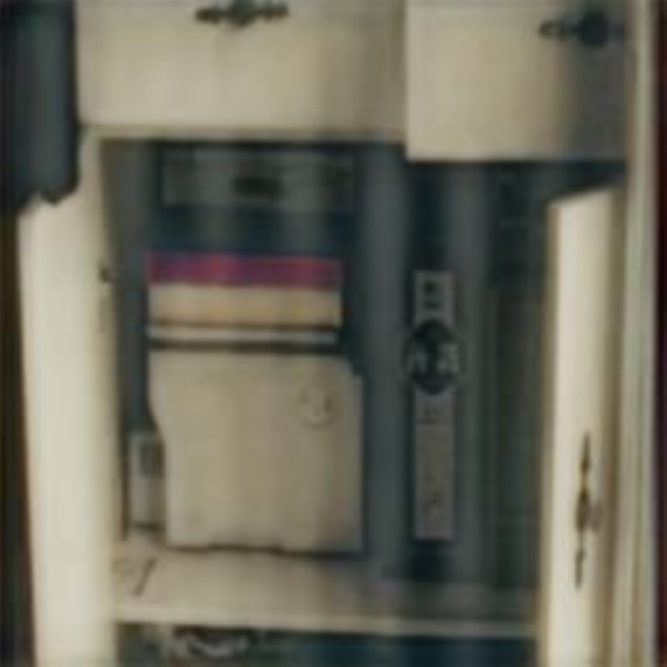}
    \end{subfigure}
    \begin{subfigure}[t]{0.13\textwidth}
    \includegraphics[width=\textwidth,height=25mm]{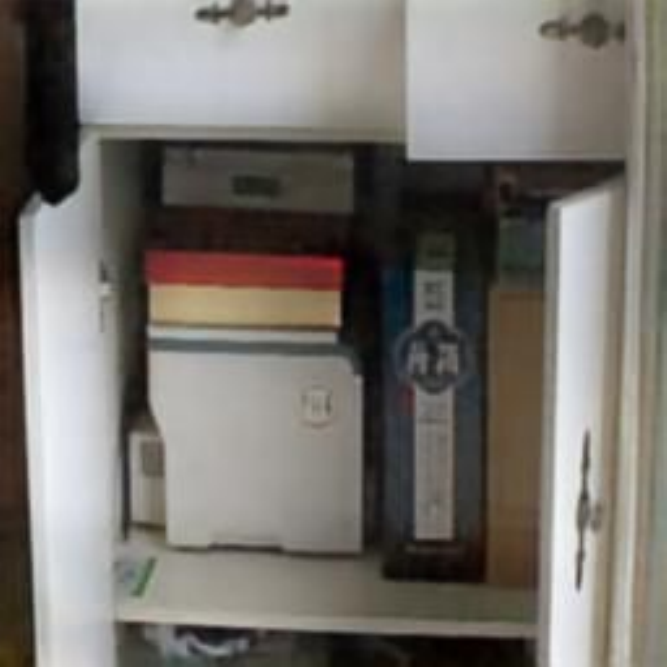}
    \end{subfigure}
    \begin{subfigure}[t]{0.13\textwidth}
    \includegraphics[width=\textwidth,height=25mm]{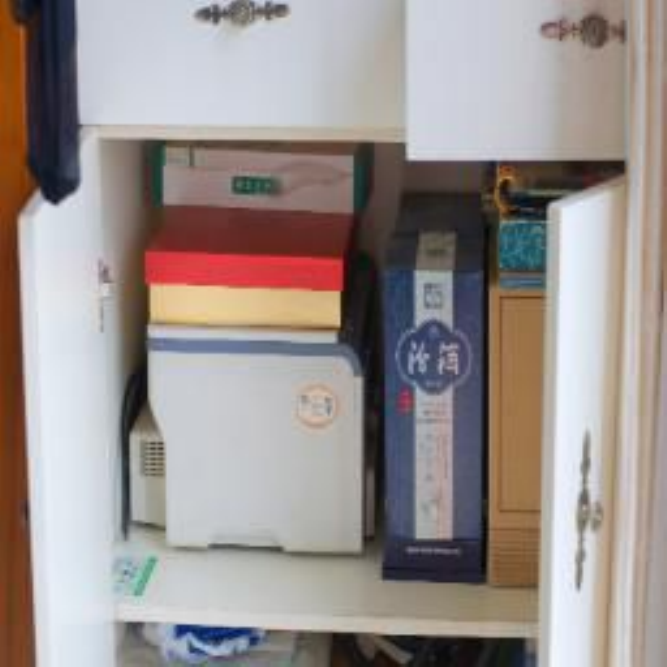}
    \end{subfigure} 
    \\
    \begin{subfigure}[t]{0.13\textwidth}
    \includegraphics[width=\textwidth,height=25mm]{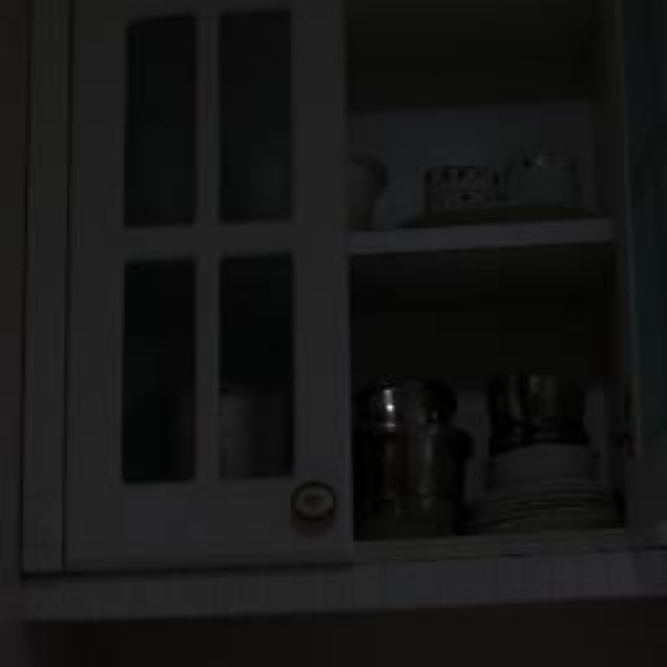}
    \end{subfigure}
    \begin{subfigure}[t]{0.13\textwidth}
    \includegraphics[width=\textwidth,height=25mm]{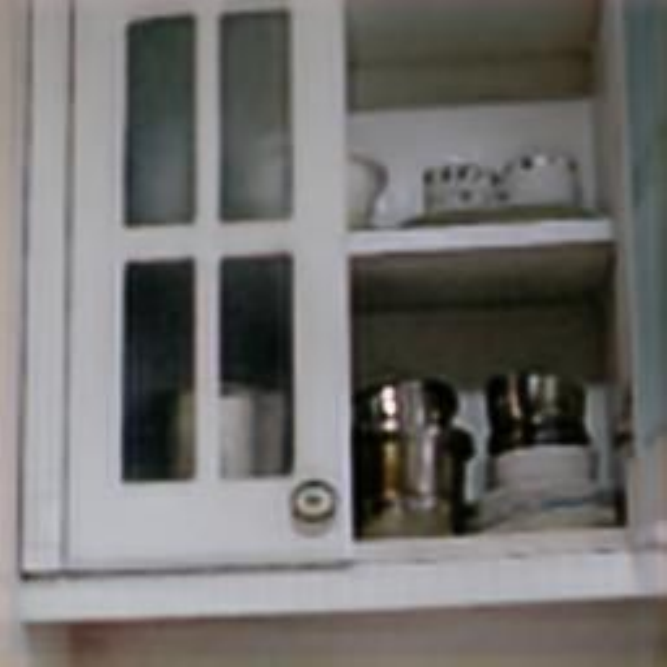}
    \end{subfigure}
    \begin{subfigure}[t]{0.13\textwidth}
    \includegraphics[width=\textwidth,height=25mm]{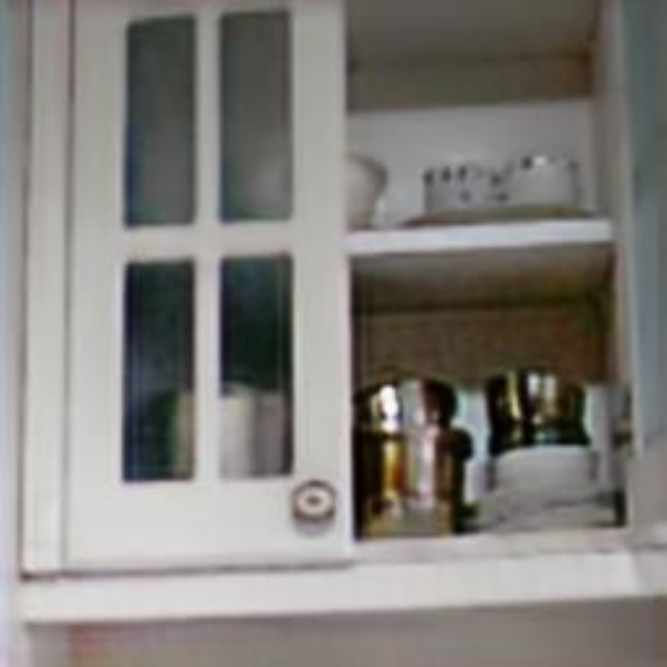}
    \end{subfigure}
    \begin{subfigure}[t]{0.13\textwidth}
    \includegraphics[width=\textwidth,height=25mm]{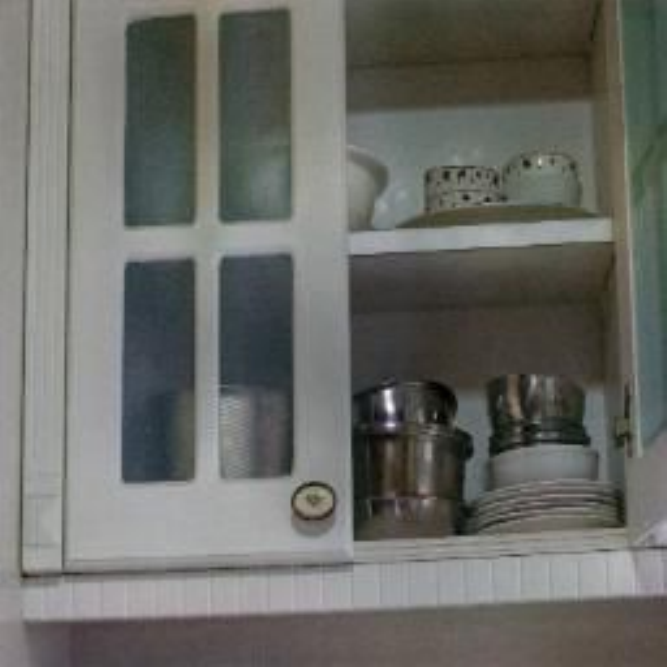}
    \end{subfigure}
    \begin{subfigure}[t]{0.13\textwidth}
    \includegraphics[width=\textwidth,height=25mm]{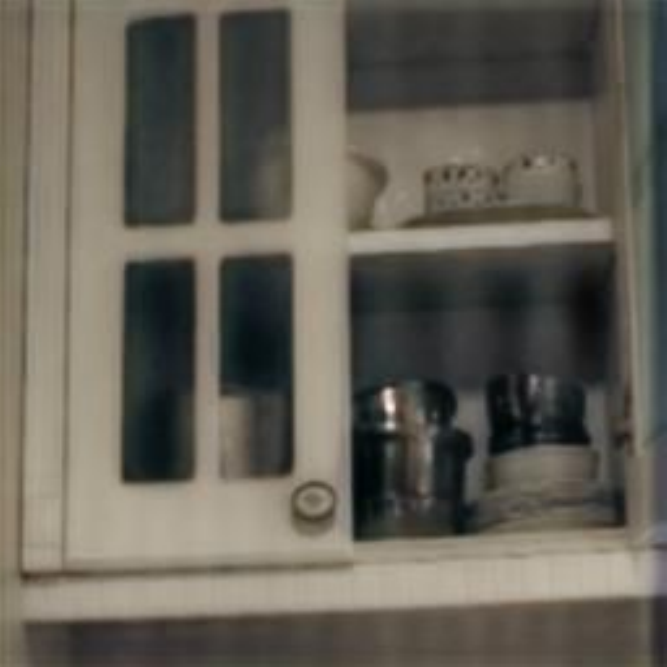}
    \end{subfigure}
    \begin{subfigure}[t]{0.13\textwidth}
    \includegraphics[width=\textwidth,height=25mm]{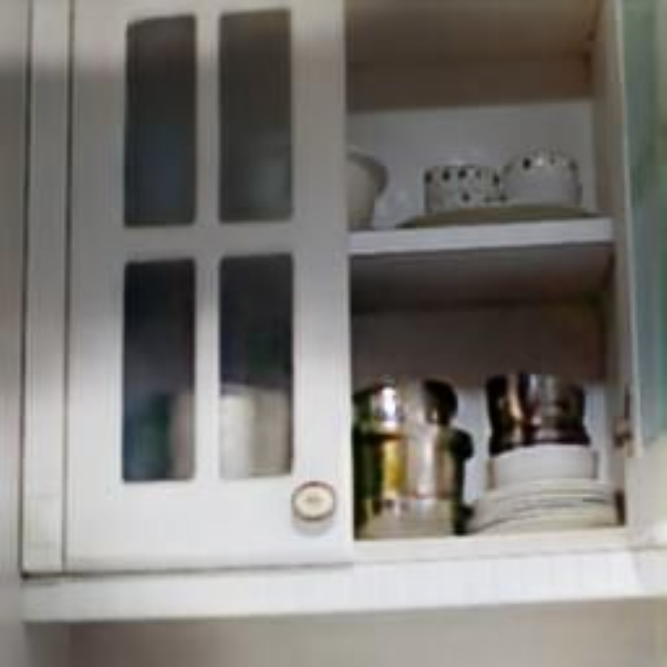}
    \end{subfigure}
    \begin{subfigure}[t]{0.13\textwidth}
    \includegraphics[width=\textwidth,height=25mm]{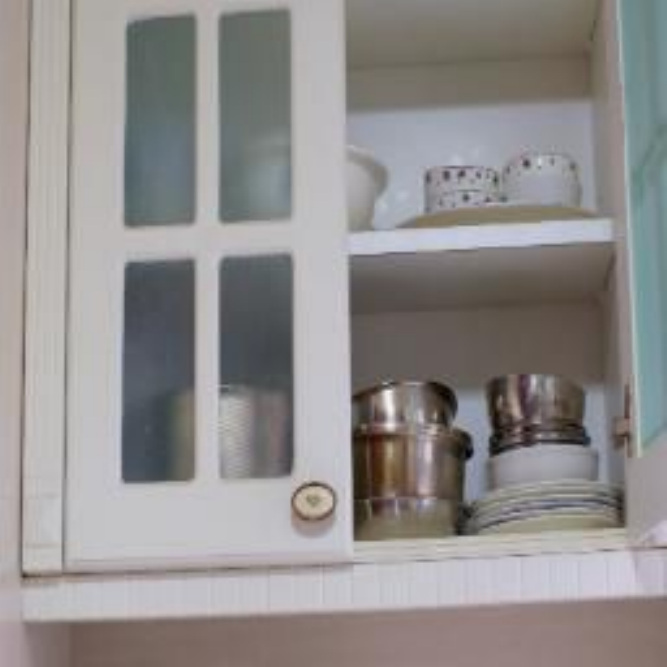}
    \end{subfigure} 
    \\
    \begin{subfigure}[t]{0.13\textwidth}
    \includegraphics[width=\textwidth,height=25mm]{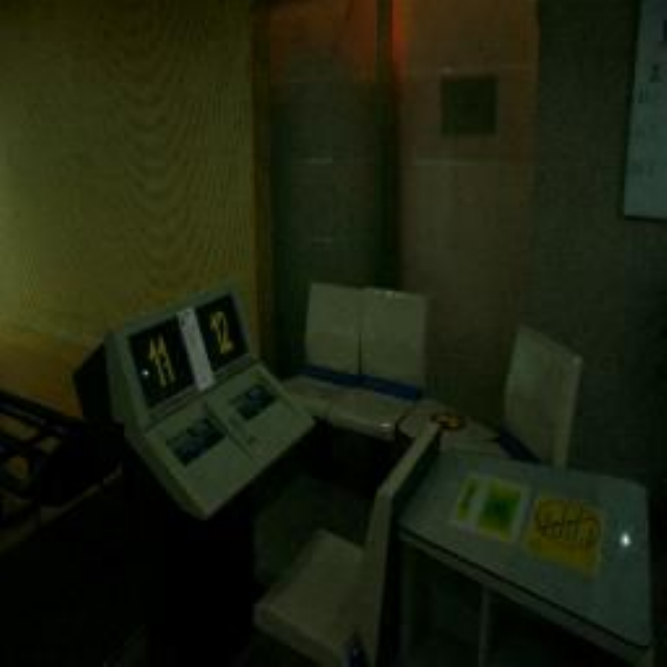}
    \end{subfigure}
    \begin{subfigure}[t]{0.13\textwidth}
    \includegraphics[width=\textwidth,height=25mm]{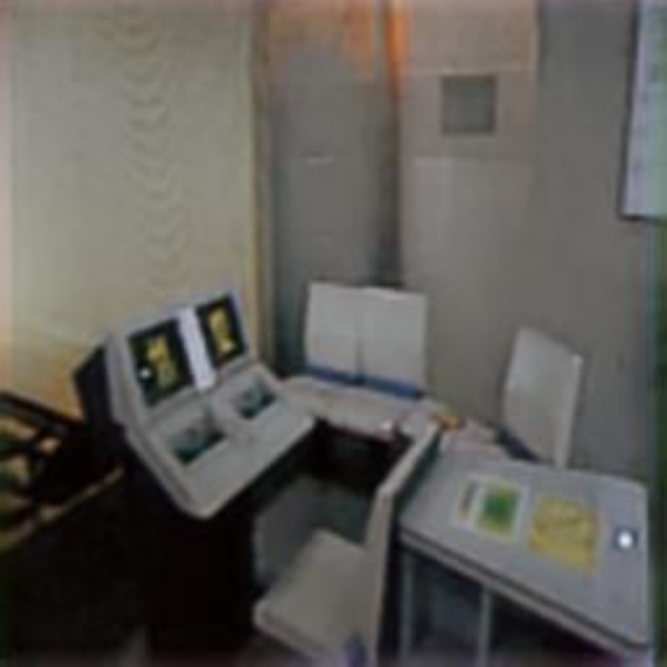}
    \end{subfigure}
    \begin{subfigure}[t]{0.13\textwidth}
    \includegraphics[width=\textwidth,height=25mm]{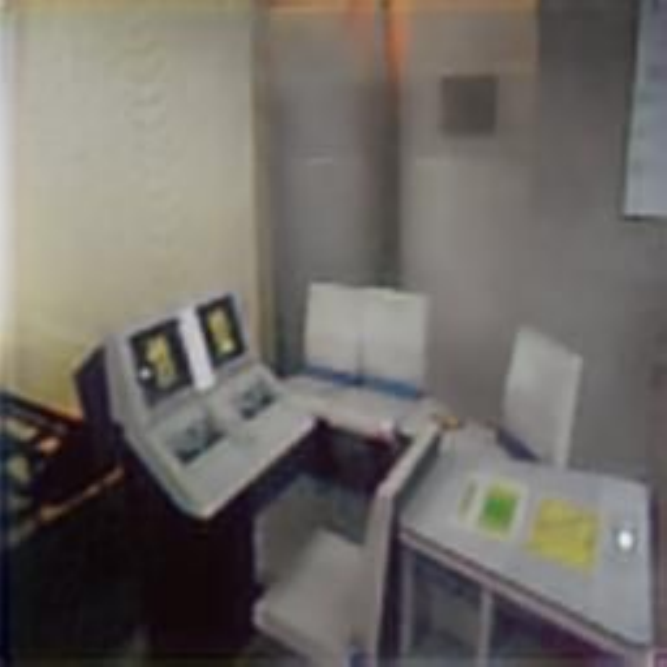}
    \end{subfigure}
    \begin{subfigure}[t]{0.13\textwidth}
    \includegraphics[width=\textwidth,height=25mm]{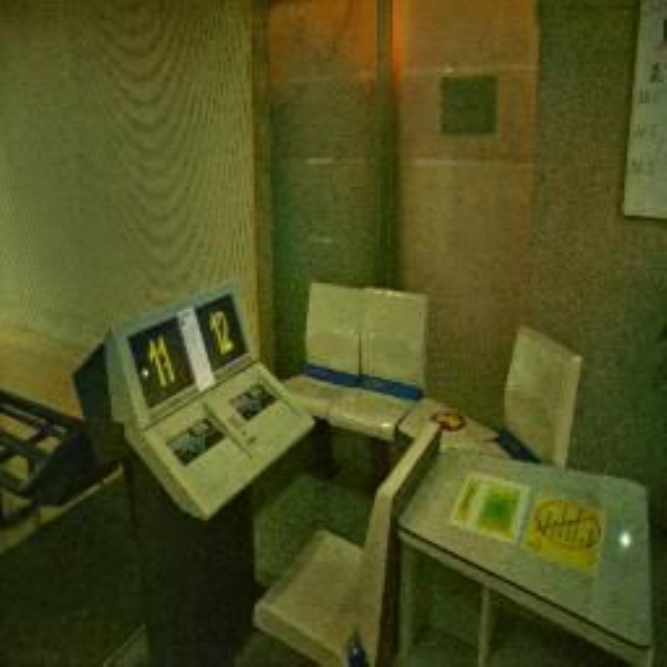}
    \end{subfigure}
    \begin{subfigure}[t]{0.13\textwidth}
    \includegraphics[width=\textwidth,height=25mm]{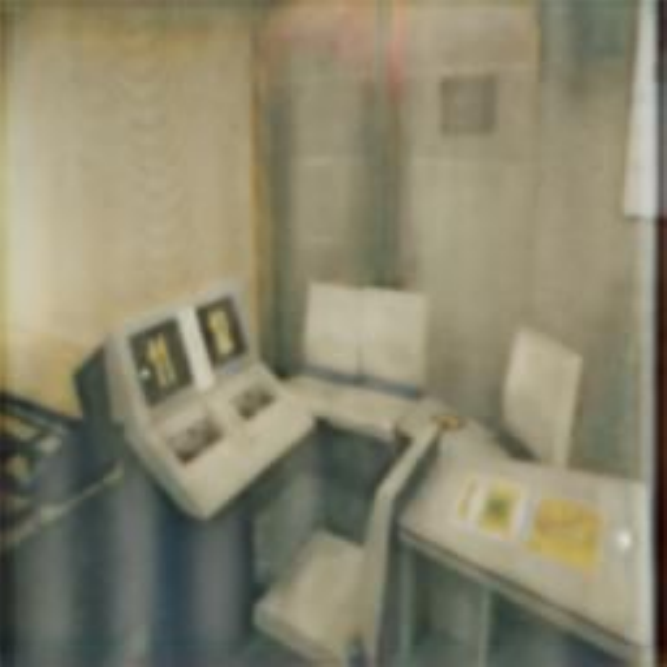}
    \end{subfigure}
    \begin{subfigure}[t]{0.13\textwidth}
    \includegraphics[width=\textwidth,height=25mm]{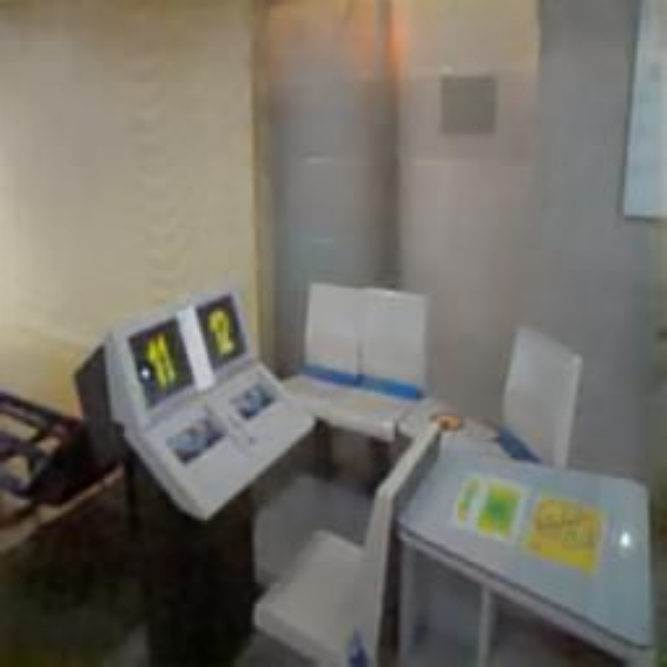}
    \end{subfigure}
    \begin{subfigure}[t]{0.13\textwidth}
    \includegraphics[width=\textwidth,height=25mm]{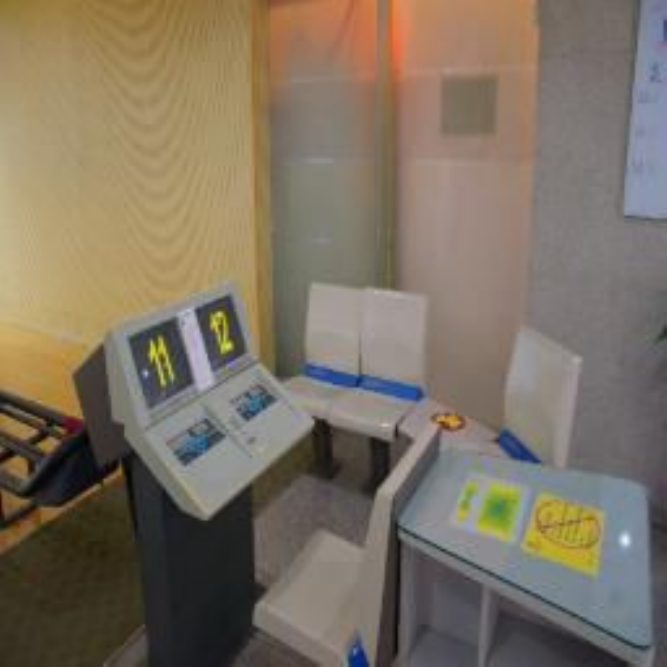}
    \end{subfigure} 
    \\
    \begin{subfigure}[t]{0.13\textwidth}
    \includegraphics[width=\textwidth,height=25mm]{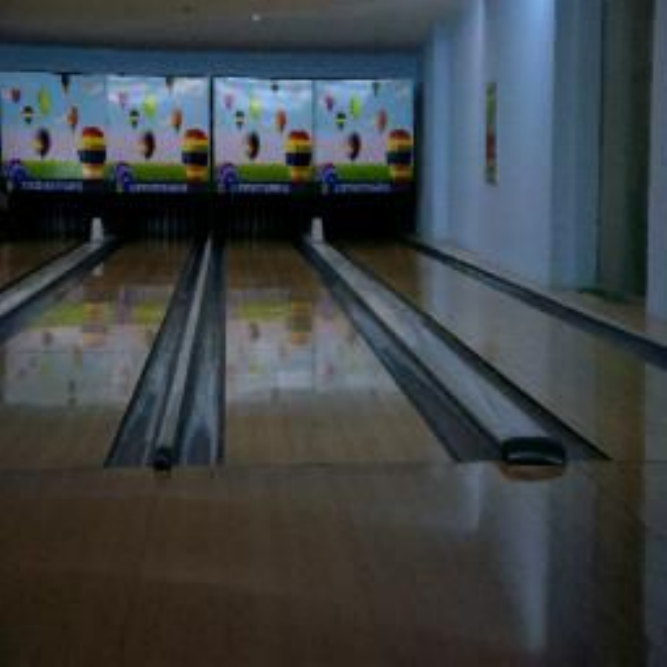}
    \end{subfigure}
    \begin{subfigure}[t]{0.13\textwidth}
    \includegraphics[width=\textwidth,height=25mm]{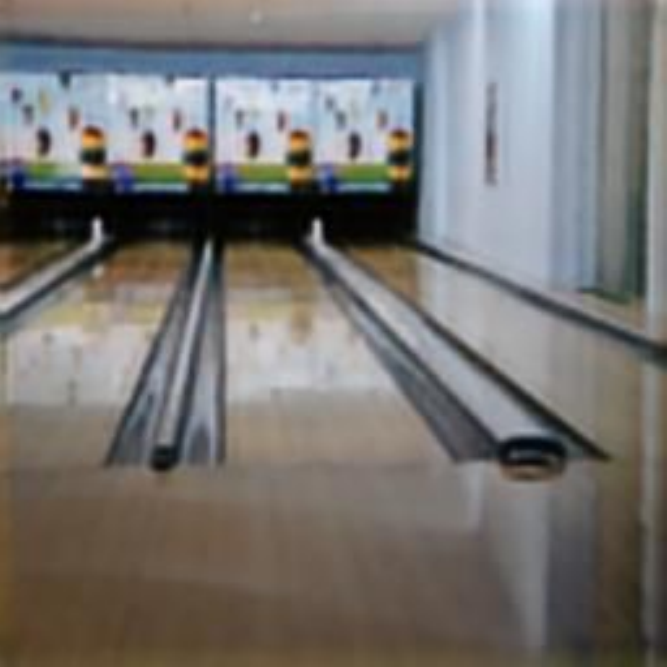}
    \end{subfigure}
    \begin{subfigure}[t]{0.13\textwidth}
    \includegraphics[width=\textwidth,height=25mm]{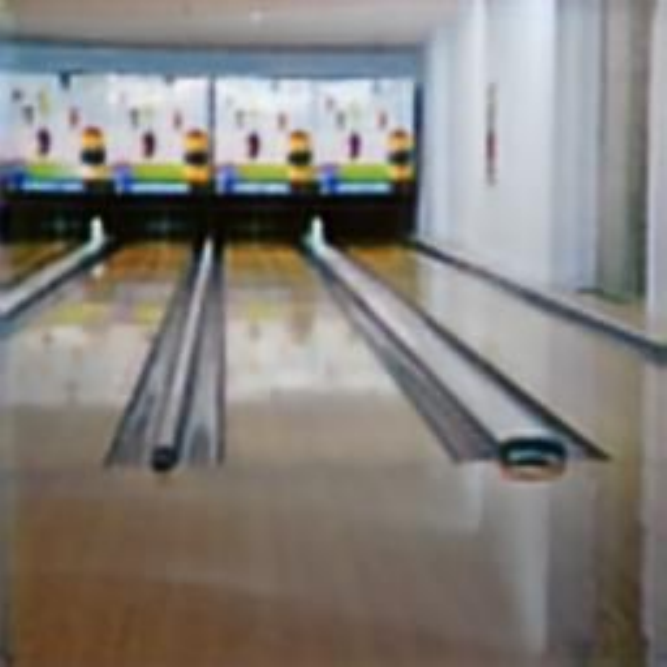}
    \end{subfigure}
    \begin{subfigure}[t]{0.13\textwidth}
    \includegraphics[width=\textwidth,height=25mm]{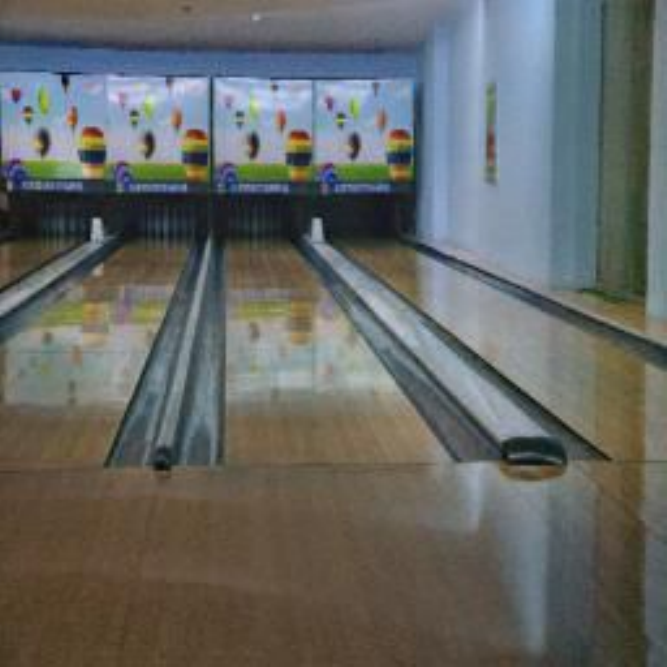}
    \end{subfigure}
    \begin{subfigure}[t]{0.13\textwidth}
    \includegraphics[width=\textwidth,height=25mm]{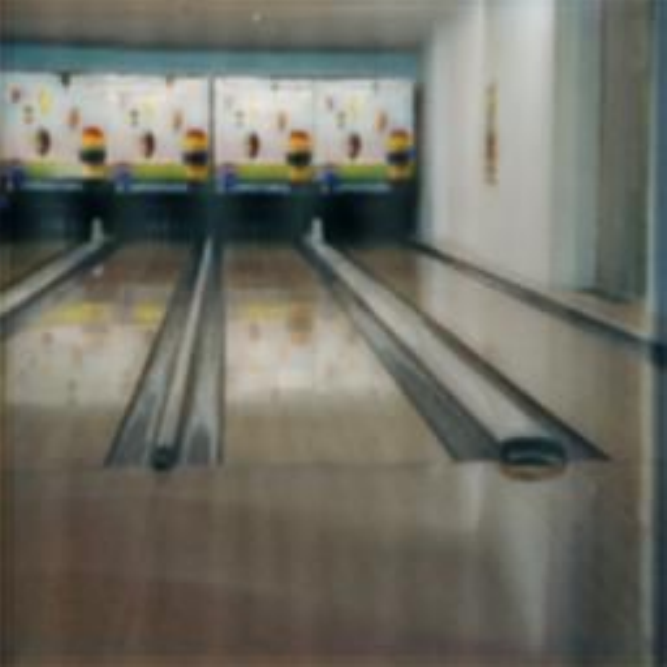}
    \end{subfigure}
    \begin{subfigure}[t]{0.13\textwidth}
    \includegraphics[width=\textwidth,height=25mm]{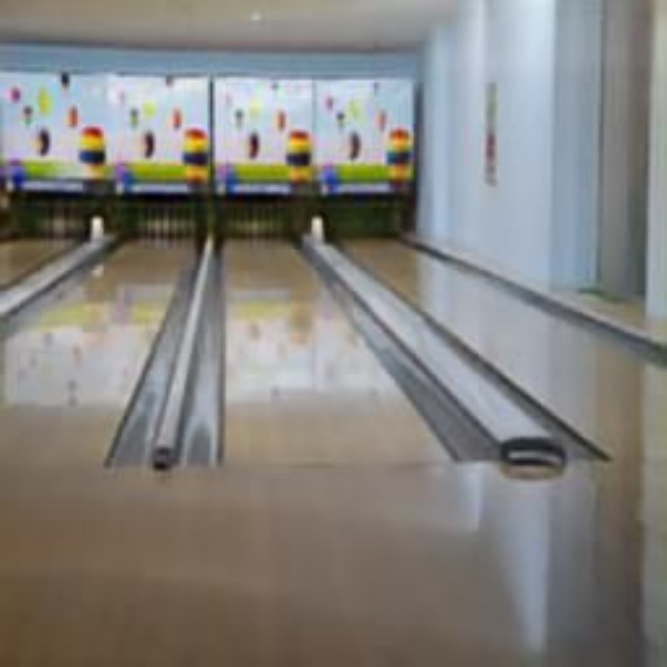}
    \end{subfigure}
    \begin{subfigure}[t]{0.13\textwidth}
    \includegraphics[width=\textwidth,height=25mm]{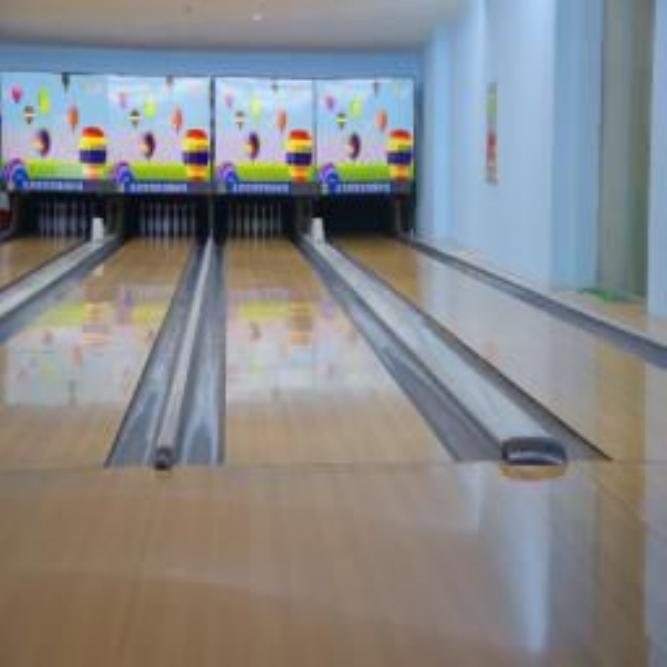}
    \end{subfigure} 
    \\
    \begin{subfigure}[t]{0.13\textwidth}
    \includegraphics[width=\textwidth,height=25mm]{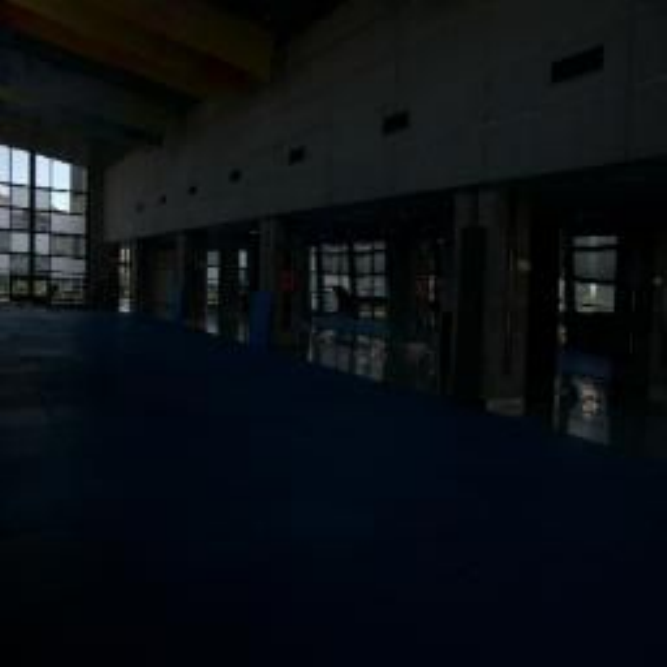}
    \end{subfigure}
    \begin{subfigure}[t]{0.13\textwidth}
    \includegraphics[width=\textwidth,height=25mm]{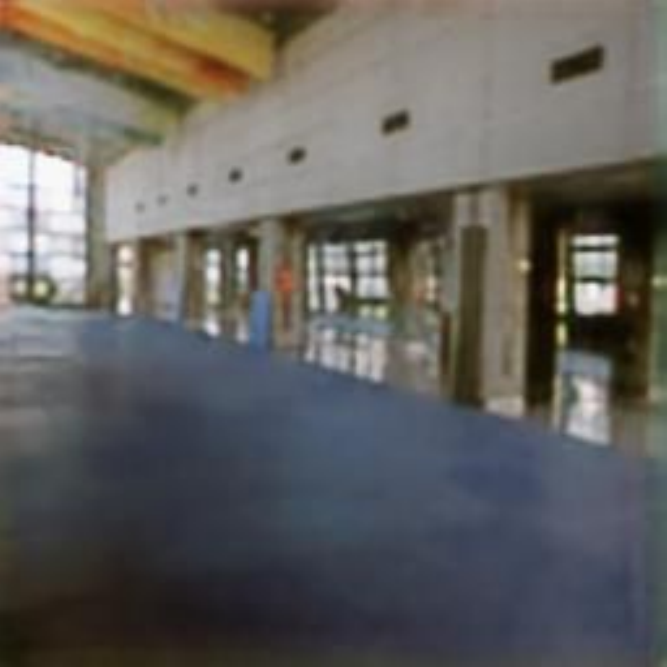}
    \end{subfigure}
    \begin{subfigure}[t]{0.13\textwidth}
    \includegraphics[width=\textwidth,height=25mm]{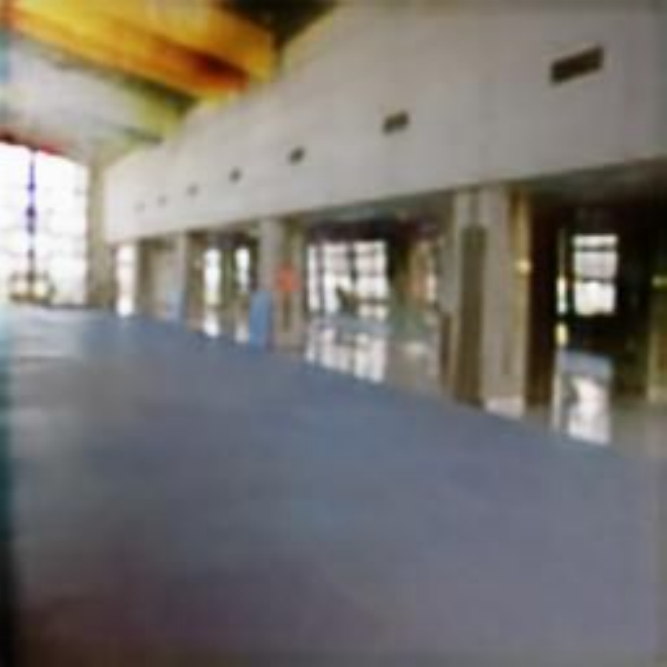}
    \end{subfigure}
    \begin{subfigure}[t]{0.13\textwidth}
    \includegraphics[width=\textwidth,height=25mm]{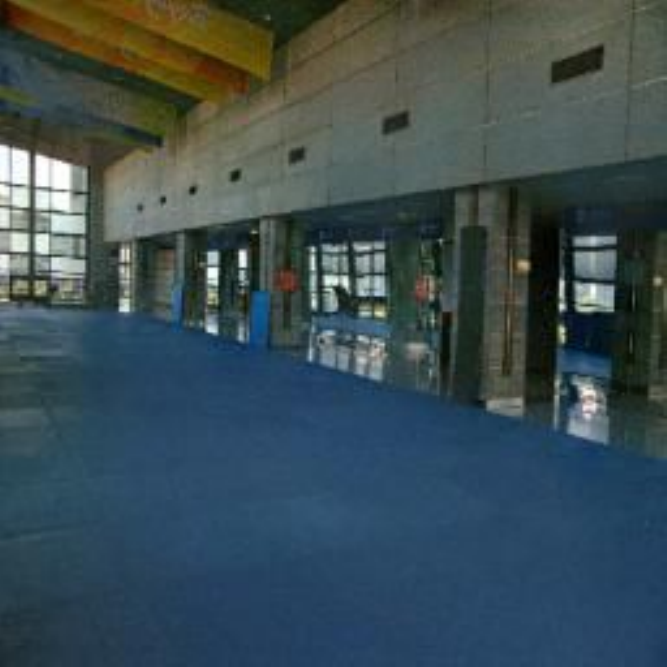}
    \end{subfigure}
    \begin{subfigure}[t]{0.13\textwidth}
    \includegraphics[width=\textwidth,height=25mm]{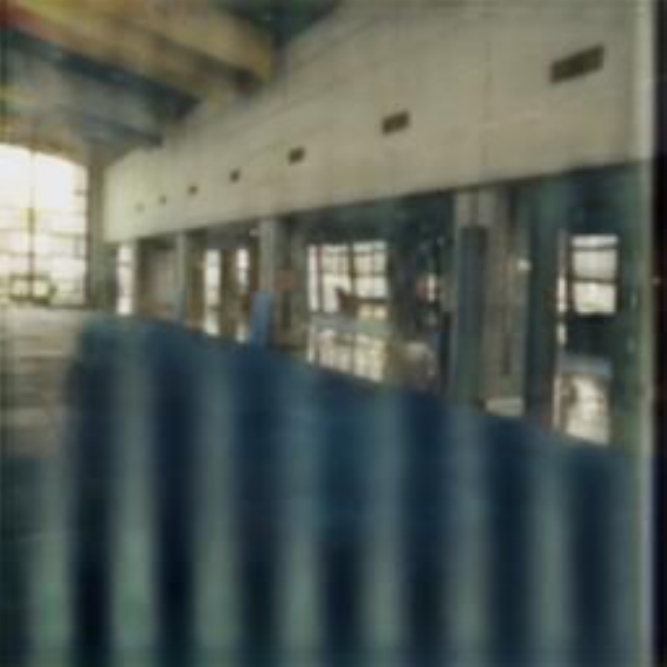}
    \end{subfigure}
    \begin{subfigure}[t]{0.13\textwidth}
    \includegraphics[width=\textwidth,height=25mm]{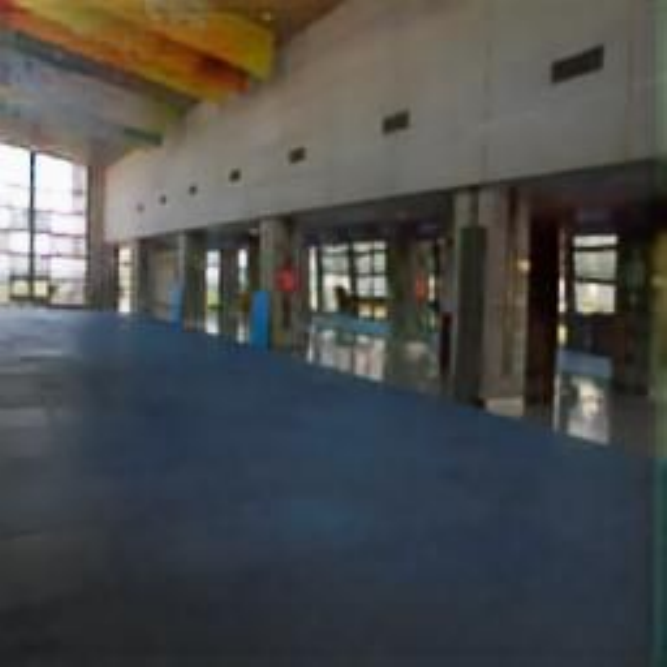}
    \end{subfigure}
    \begin{subfigure}[t]{0.13\textwidth}
    \includegraphics[width=\textwidth,height=25mm]{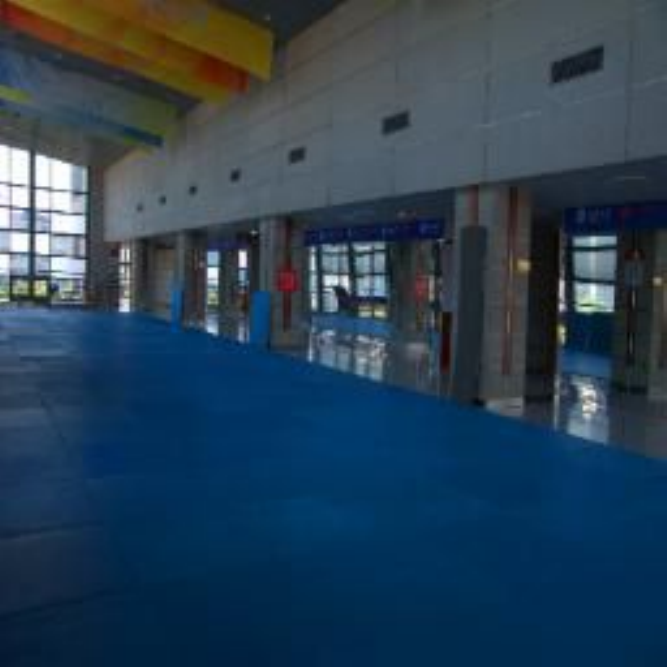}
    \end{subfigure} 
    \caption{Qualitative comparison with different models.} 
    \label{qual}
\end{figure*}

\begin{figure*}
    \centering
    \frame{\includegraphics[width=\linewidth]{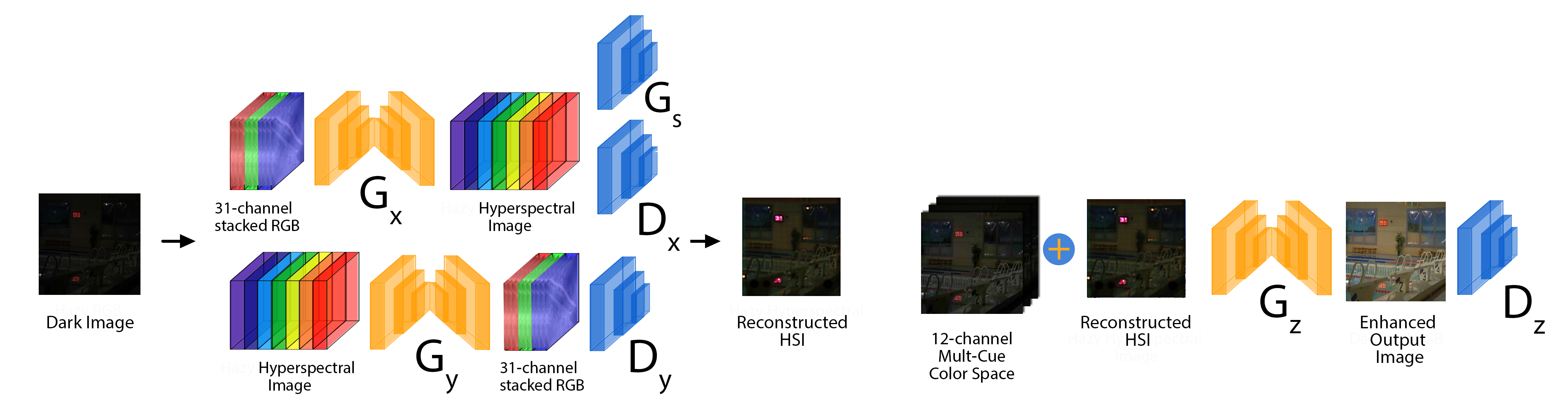}}
    \caption{The schematic diagram for the proposed SpecNet}
    \label{fig:overview}
\end{figure*}

\begin{figure*}[t]
    \centering
    \includegraphics[width=\linewidth]{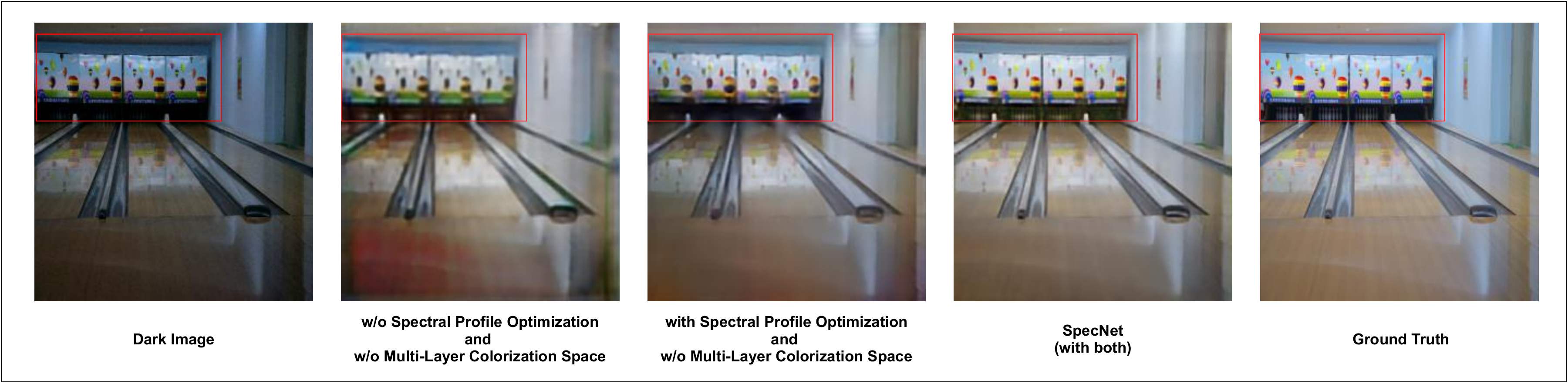}
    \caption{Qualitative comparison for ablation models}
    \label{ablation_img}
\end{figure*}

\section{Supplementary Material}
\subsection{Architectural Details}
The proposed work adapts an unpaired cycle-consistency framework ~\cite{zhu2017unpaired} to exploit supervision at the level of sets.
The objective is to learn a  mapping function $G_x: X_{31} \,\to\, Y$, where $X_{31}$ represents the stacked RGB image and $Y$ refers to the reconstructed HSI. In context to the adversarial loss, the reconstruction module can be expressed as

\begin{align}
     \begin{split}
        &\Lagr_{GAN_x}(G_x,D_x, X_{31}, Y) \\
        &= E_{x_{31}\sim p_{data(x_{31})}}[log(1-D_x(G_x(x_{31})))] \\
&+ E_{y\sim p_{data(y)}}[log(D_x(y))] \\
    \end{split}
    \\
    \begin{split}
        &\Lagr_{GAN_y}(G_y,D_y, Y, X_{31}) \\
        &=  E_{y\sim p_{data(y)}}[log(1-D_y(G_y(y)))] \\
&+E_{x_{31}\sim p_{data(x_{31})}}[log(D_y(x_{31}))] \\
    \end{split}
    \\
    \begin{split}
        \Lagr_{GAN} = \Lagr_{GAN_x} + \Lagr_{GAN_y}
    \end{split}
\end{align}

Figure \ref{fig:overview} shows the complete schematic diagram of the proposed SpecNet.
The generators in $G_x$, $G_y$ and $G_z$ adopt a U-Net with skip connections while PatchGAN is adopted for the corresponding discriminators. We use L1 cycle consistency losses and identity losses~\cite{zhu2017unpaired} to further improve the reconstructed HSI.

The generator $G_s$ uses a ResNet-based architecture to compute the spectral profile of input image. Deriving inspiration from recent work by \citeauthor{durall2020watch}, we extend the analysis to hyperspectral images. The network aims to regularize the generated HSI with respect to spectral distribution of real images.

\section{Datasets}
To facilitate HSI reconstruction, HSCycle is trained using ICVL BGU Hyperspectral Dataset (NTIRE 2018) ~\cite{arad2016sparse, arad2018ntire} and the NTIRE 2020 dataset. The dataset is composed of 200 natural images with various indoor and outdoor scenes. The dataset provides sampled images which each having 31 spectral bands. Adjacent bands have an incremental difference of 10 nm. In addition, preprocessing like random cropping and flip is utilized to increase the total number of images upto 6000.

To train the proposed network for low light image enhancement, we use low/normal-light pairs in the LOL Dataset. The LOL Dataset consists of 500 image pairs, which is pre-divided into training and evaluation datasets.

\subsection{Additional Results}
In Figure \ref{qual}, we show additional qualitative comparison of \textsc{SpecNet} with several deep learning based models. U-Net, Pix2Pix and CycleGAN, being general computer vision models, were re-trained on the train dataset used by \textsc{SpecNet}. 

In Figure \ref{ablation_img} we visually show the performance of \textsc{SpecNet} with respect to other ablated models. The red box highlights the improvement our model gets due Spectral Profile optimization and multi-layer colorization. 
\section{Acknowledgements}
This work is supported by BITS Additional Competitive
Research Grant (PLN/AD/2018-19/5).
\bibliography{sample}

\begin{thebibliography}{9}
\providecommand{\natexlab}[1]{#1}
\providecommand{\url}[1]{\texttt{#1}}
\providecommand{\urlprefix}{URL }
\expandafter\ifx\csname urlstyle\endcsname\relax
  \providecommand{\doi}[1]{doi:\discretionary{}{}{}#1}\else
  \providecommand{\doi}{doi:\discretionary{}{}{}\begingroup
  \urlstyle{rm}\Url}\fi

\bibitem[{Arad and Ben-Shahar(2016)}]{arad2016sparse}
Arad, B.; and Ben-Shahar, O. 2016.
\newblock Sparse recovery of hyperspectral signal from natural RGB images.
\newblock In \emph{European Conference on Computer Vision}, 19--34. Springer.

\bibitem[{Arad, Ben-Shahar, and Timofte(2018)}]{arad2018ntire}
Arad, B.; Ben-Shahar, O.; and Timofte, R. 2018.
\newblock Ntire 2018 challenge on spectral reconstruction from rgb images.
\newblock In \emph{Proceedings of the IEEE Conference on Computer Vision and
  Pattern Recognition Workshops}, 929--938.

\bibitem[{Durall, Keuper, and Keuper(2020)}]{durall2020watch}
Durall, R.; Keuper, M.; and Keuper, J. 2020.
\newblock Watch your Up-Convolution: CNN Based Generative Deep Neural Networks
  are Failing to Reproduce Spectral Distributions.
\newblock In \emph{Proceedings of the IEEE/CVF Conference on Computer Vision
  and Pattern Recognition}, 7890--7899.

\bibitem[{Isola et~al.(2017)Isola, Zhu, Zhou, and Efros}]{isola2017image}
Isola, P.; Zhu, J.-Y.; Zhou, T.; and Efros, A.~A. 2017.
\newblock Image-to-image translation with conditional adversarial networks.
\newblock In \emph{Proceedings of the IEEE conference on computer vision and
  pattern recognition}, 1125--1134.

\bibitem[{Jiang et~al.(2019)Jiang, Gong, Liu, Cheng, Fang, Shen, Yang, Zhou,
  and Wang}]{jiang2019enlightengan}
Jiang, Y.; Gong, X.; Liu, D.; Cheng, Y.; Fang, C.; Shen, X.; Yang, J.; Zhou,
  P.; and Wang, Z. 2019.
\newblock Enlightengan: Deep light enhancement without paired supervision.
\newblock \emph{arXiv preprint arXiv:1906.06972} .

\bibitem[{Mehta et~al.(2020)Mehta, Sinha, Mandal, and
  Narang}]{mehta2020domainaware}
Mehta, A.; Sinha, H.; Mandal, M.; and Narang, P. 2020.
\newblock Domain-Aware Unsupervised Hyperspectral Reconstruction for Aerial
  Image Dehazing.
\newblock \emph{arXiv:2011.03677} .

\bibitem[{Ronneberger, Fischer, and Brox(2015)}]{ronneberger2015u}
Ronneberger, O.; Fischer, P.; and Brox, T. 2015.
\newblock U-net: Convolutional networks for biomedical image segmentation.
\newblock In \emph{International Conference on Medical image computing and
  computer-assisted intervention}, 234--241. Springer.

\bibitem[{Wei et~al.(2018)Wei, Wang, Yang, and Liu}]{wei2018deep}
Wei, C.; Wang, W.; Yang, W.; and Liu, J. 2018.
\newblock Deep retinex decomposition for low-light enhancement.
\newblock In \emph{British Machine Vision Conference}.

\bibitem[{Zhu et~al.(2017)Zhu, Park, Isola, and Efros}]{zhu2017unpaired}
Zhu, J.-Y.; Park, T.; Isola, P.; and Efros, A.~A. 2017.
\newblock Unpaired image-to-image translation using cycle-consistent
  adversarial networks.
\newblock In \emph{Proceedings of the IEEE international conference on computer
  vision}, 2223--2232.

\end{thebibliography}
\end{document}